\documentclass[12pt,preprint]{aastex}

\newcommand{\Msun}{$M_{\sun}$}

\newcommand{\sn}{$S_{\rm N}$}

\newcommand{\tv}{$T(V)$}
\newcommand{\tmp}{$T_{\rm MP}$}
\newcommand{\fmp}{$F_{\rm MP}$}
\newcommand{\tm}{$F(V)$}
\newcommand{\hubble}{{\it Hubble Space Telescope\/}} 
\newcommand{\hst}{{\it HST\/}} 

\newcommand\cola {\null}
\newcommand\colb {&}
\newcommand\colc {&}
\newcommand\cold {&}
\newcommand\cole {&}
\newcommand\colf {&}
\newcommand\colg {&}
\newcommand\colh {&}

\newcommand\eol{\\}

\begin{document}

\title{The Globular Cluster Luminosity Function and Specific
Frequency in Dwarf Elliptical Galaxies\footnote{Based on
observations with the NASA/ESA {\it Hubble Space Telescope}, obtained
at the Space Telescope Science Institute, which is operated by the
Association of Universities for Research in Astronomy, Inc., under
NASA contract No.NAS5-26555}}

\author{Bryan W. Miller} 
\affil{Gemini Observatory, Casilla 603, La Serena, Chile}
\author{Jennifer M. Lotz\footnote{NOAO Leo Goldberg Fellow}} 
\affil{National Optical Astronomy Observatory, 950 N. Cherry Ave.,
  Tucson, AZ 85719, USA}

\begin{abstract}
The globular cluster luminosity function, specific globular cluster
frequency, \sn, specific globular cluster mass, \tmp, and globular
cluster mass fraction in dwarf elliptical galaxies are explored using
the full 69 galaxy sample of the \hst\ WFPC2 Dwarf Elliptical Galaxy
Snapshot Survey.  The GCLFs of the dEs are well-represented with a
$t_5$ function with a peak at $M_{V,Z}^0(dE,HST) = -7.3 \pm 0.1$.
This is $\sim0.3$ magnitudes fainter than the GCLF peaks in giant
spiral and elliptical galaxies, but the results are consistent within
the uncertainties.  The bright-end slope of the luminosity
distribution has a power-law form with slope $\alpha = -1.9\pm0.1$.
The trend of increasing \sn\ or \tmp\ with decreasing host galaxy
luminosity is confirmed.  The mean value for \tmp\ in dE,N galaxies is
about a factor of two higher than the mean value for non-nucleated
galaxies and the distributions of \tmp\ in dE,N and dE,noN galaxies
are statistically different. These data are combined with results from
the literature for a wide range of galaxy types and environments.  At
low host galaxy masses the distribution of \tmp\ for dE,noN and dI
galaxies are similar.  This supports the idea that one pathway for
forming dE,noN galaxies is by the stripping of dIs.  The formation of
nuclei and the larger values of \tmp\ in dE,N galaxies may be due to
higher star formation rates and star cluster formation efficiencies
due to interactions in galaxy cluster environments.
\end{abstract}


\section{Introduction}

The properties of the globular cluster (GC) systems of galaxies
contain important information about the formation and evolution of the
star clusters themselves as well as their host galaxies.  In the local
universe massive, compact young star clusters that are likely
progenitors of traditionally old GCs are usually found in starburst
environments such as galaxy mergers
\citep[e.g.][]{holtz92,mil97,whit99} or star bursting dwarfs
\citep[e.g.][]{billett02,alonso04,dgwt05}.  The metallicity of the gas
out of which they form and the time since their formation can be
gleaned from the properties of their simple stellar populations.
Therefore, massive star clusters are good tracers of significant star
forming episodes in their host galaxies.  Also, the oldest GCs in the
Milky Way have ages consistent with a Hubble time, suggesting that
they were among the earliest stellar systems to form and therefore
provide us with a window into conditions in the very early universe.

The observed star cluster systems in nearby galaxies are the result of
the interplay between the conditions and processes during cluster
formation and the processes that destroy clusters.  The number of
clusters that form in a given potential well depends on the cluster
formation efficiency, the cluster formation history, the detailed
physics of the cluster formation process, and the initial cluster mass
function.  All of these factors may be dependent on the initial
environmental conditions such as the amount of available gas, the
metallicity of the gas, and the external pressure
\citep{fr85,mp96,ee97}.  Star clusters can then be destroyed by a
combination of stellar evolution, two-body relaxation, and tidal
shocking \citep[e.g.][]{go97,baum98,fz01,vz03}.  In addition, cluster
systems can be both augmented by the accretion of clusters during
interactions or mergers and reduced by stripping via strong external
tides \citep{cote98,beasley02}.  Only by studying the properties of
star clusters in different environments where different sets of
processes may be significant can we constrain our theories of star
cluster formation and evolution.

Dwarf galaxies are particularly useful objects for studies of massive
star clusters.  A census of the globular clusters in Local Group dEs
showed that there is a large scatter in the number per galaxy.
However, those that do form clusters often have specific globular
cluster frequencies (\sn, number per unit luminosity) as high as those
in giant elliptical galaxies \citep{harris91}. Larger surveys of dEs
in the Virgo and Fornax Clusters revealed a trend of increasing \sn,
or an increase in the scatter of \sn, with decreasing galaxy
luminosity \citep{durrell96,mil98}.  This trend is consistent with a
constant GC formation efficiency but decreasing field star formation
efficiency because of supernovae feedback from the forming clusters.
\cite{mil98} also showed that the mean \sn\ for nucleated dEs is about
double that for non-nucleated dEs.  This has implications for the
formation of nuclei and the progenitor galaxies of non-nucleated
versus nucleated dEs.  However, the sample size was still relatively
small (24 dEs in Miller et al.'s HST sample) and the comparison with
the GC systems of dIs was limited to dIs in the Local Group, which is
a very different environment from the Virgo or Fornax Clusters.
Therefore, the \hst\ WFPC2 surveys were extended to include more dE
galaxies and some dI galaxies in the Virgo and Fornax Clusters
\citep{lmf04,seth04}. Also, a search for GCs in 57 dwarfs in nearby
groups and in the field was carried out by \cite{sharina05} and the
Virgo and Fornax Clusters have also been surveyed with ACS on HST
\citep{cote04,jordan07a}.  Finally the sample of galaxies with studied
cluster systems is large enough to start making meaningful comparisons
between different types of dwarfs in different environments.

These large samples are also needed for studying the globular cluster
luminosity function (GCLF) in dEs.  Each dwarf has only a few GCs, so
samples from different galaxies must be combined into a composite GCLF
\citep{durrell96}. Because dwarfs have lower density disks than
galaxies like the Milky Way, the dominant processes that destroy GC
will work more slowly in them.  Therefore, the GCLF, more precisely
the GC mass function, in dwarfs has presumably undergone less
modification and is closer to the initial GC mass function than in the
GC systems of giant spiral or elliptical galaxies.  Any differences in
the GC mass function must be explainable by theories of GC formation
and destruction combined with hierarchical models of the merging of
galaxies and the formation of GC systems
\citep[e.g.][]{beasley02,kravtsov05}.

Therefore, in this paper we present an analysis of the GCLF and
specific luminosities and masses for the full 69 galaxy sample from
the \hst\ WFPC2 Dwarf Elliptical Snapshot Survey
\citep{mil98,stiavelli01,lotz01,lmf04}.  The observations are
summarized in Section~\ref{sec:obs} and the new analysis and results
are given in Section~\ref{sec:results} and discussed in
Section~\ref{sec:disc}.  Throughout the work use the \hst\ Cepheid Key
Project \citep{freedman01} distances to the galaxy clusters:
$(m-M)_0({\rm Leo}) = 30.0$; $(m-M)_0({\rm Virgo}) = 30.92$;
$(m-M)_0({\rm Fornax}) = 31.39$.

\section{Observations}\label{sec:obs}

Imaging of the 69 dE galaxies in the \hst\ Dwarf Elliptical Snapshot
Survey was obtained with WFPC2 on the \hubble\ in Cycles 6, 7, and 9
(GO programs 6352, 7377, and 8500).  The galaxies were selected from
catalogs of the Virgo and Fornax Clusters \citep{bst85,f89} and the
Leo Group \citep{fs90}.  The dE galaxies have absolute magnitudes in
the range $-12 < M_B < -17$ and includes 45 nucleated galaxies and 24
non-nucleated galaxies.  The images were taken with the F555W ($2
\times 230$~sec) and F814W (300~sec) filters and the galaxies were
centered on chip WF3 to maximize the spatial coverage.  More details
about the final data reduction, the selection of the GC candidates,
and the photometry can be found in \citet{lmf04}.

The global properties of the sample galaxies and the net number of GC
candidates per galaxy are given in Tables~\ref{tab:densmpl} and
\ref{tab:denonsmpl} for nucleated and non-nucleated galaxies,
respectively.  To quickly summarize \cite{lmf04}, the GC candidates are
selected to be compact (${\rm FHWM_{F555W}} < 2.5$~pixels) and to have
colors typical of GCs ($0.5 < V-I < 1.5$ with $\sigma(V-I) < 0.3$).
``Object+background'' candidates are those within 5 scale lengths of
the centers of the galaxies while ``background'' candidates are those
detected outside of 5 scale lengths.  The net number of GC candidates
in a galaxy, $N_{\rm GC}$, is then the number of ``object+background''
candidates minus the number of ``background'' candidates scaled to the
area within 5 scale lengths. 

These values of $N_{\rm GC}$ do include the nuclei.  This is justified
for the fainter nuclei since their properties are essentially
indistinguishable from bright GCs \citep{lmf04}.  The brighter nuclei
may form from mergers of GCs or star formation on top of an existing
central GC \citep[see Sec.~\ref{sec:disc} and][]{bekki06}.  Therefore,
it is reasonable to count a bright nucleus as at least one GC.  It
should be noted that the subtraction of the bright GCs, or even all
nuclei, from $N_{\rm GC}$ does not qualitatively change the results of
this paper.

The values of $N_{\rm GC}$ in Tables~\ref{tab:densmpl} and
\ref{tab:denonsmpl} should be the same as those in Tables~2 and 3 of
\cite{lmf04}.  However, we have found some sorting errors in the last
column of Tables~2 and 3 of that paper have been corrected in the
current tables.  Only column $N_{\rm GC}$ of the tables in
\cite{lmf04} has sorting errors and this does not affect any results
in that paper.

\section{Results}\label{sec:results}

\subsection{Globular Cluster Luminosity Function}

As can been seen from Tables~\ref{tab:densmpl} and
\ref{tab:denonsmpl}, the typical galaxy in our sample has about 6 GC
candidates and the maximum number in any one galaxy is 52 in
VCC~940. Therefore, no single galaxy has enough clusters to yield an
accurate luminosity function.  Therefore, as is common practice in
these types of analyzes, we must combine the GC candidates from many
galaxies to derive a mean dE GC luminosity function (GCLF).  In this
case, we derive the GCLF for the galaxies in the Virgo and Fornax
Clusters separately since all the galaxies in each cluster are at
approximately the same distance.  Obtaining the net GCLF requires
correction for foreground/background objects and the photometric
completeness function. 

Subtracting the foreground and background contamination is similar to
what is done to determine $N_{\rm GC}$, but now all GC candidates for
a given galaxy cluster are considered as an ensemble.  The total
sample of ``object'' candidates is a merged list of all the object
candidates (within 5 scale lengths).  Likewise, the total sample of
``background'' candidates includes all candidates classified
previously as background.  The background LF is then scaled by the
ratio of the total areas used for selecting the ``object'' and
``background'' candidates.

\begin{figure}
\includegraphics[width=\textwidth]{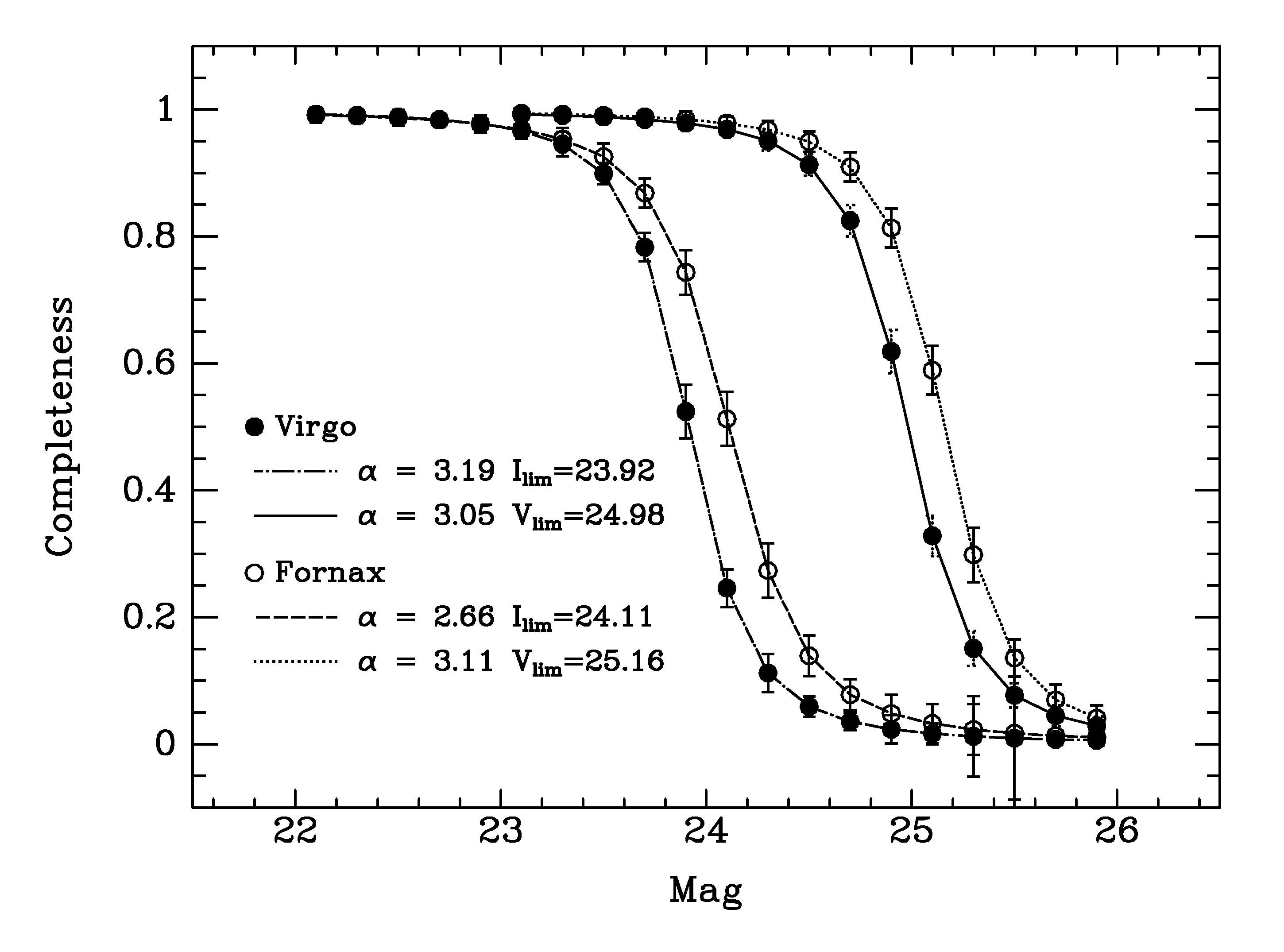}
\caption{The completeness functions of WFPC2 images in the \hst\ dE
  Snapshot Survey.  The limiting magnitudes, $V_{lim}$ and $I_{lim}$,
  where the completeness fraction drops to 0.5, differs slightly been
  the Virgo and Fornax Cluster samples.  The parameter $\alpha$ from
  Pritchet's interpolation function \citep{flemming95} is a measure of
  the sharpness of the drop in the completeness function.}
\label{fig:cplt}
\end{figure}

The completeness curve is derived using standard artificial
``cluster'' tests.  The object PSFs used for the tests are the sum of
several bright GC candidates taken from the WF3 image of
VCC~1254. Artificial GCs of different magnitudes are then added to
representative WFPC2 images using the DAOPHOT task ADDSTAR in
IRAF\footnote{IRAF is distributed by the National Optical Astronomy
Observatories, which are operated by the Association of Universities
for Research in Astronomy, Inc., under cooperative agreement with the
National Science Foundation.}.  During each trial the number of
artificial objects added is only 20\% of the number of actual objects
found in a given magnitude bin so that the artificial objects do not
affect the crowding in the image.  Hundreds of trials are run in order
to build up adequate statistics.  The same object detection algorithm
used to select the actual candidates is then used to detect the
artificial GCs and the ratio of the objects found to the objects added
gives the completeness fraction.  The completeness function is then
characterized by Pritchet's interpolation function \citep{flemming95}.
The completeness functions for the Virgo and Fornax samples are shown
in Figure~\ref{fig:cplt}.  The limiting magnitudes, where the
completeness fraction drops to 0.5, are $V_{lim} \approx 25$ and
$I_{lim} \approx 24$ but differ slightly between Virgo and Fornax
Cluster images.  This may be due to small changes in the mean sky
background.

The background subtraction and fitting of the GCLFs are done using the
method of \cite{sh93}.  The best-fitting parameters of either a
Gaussian or a $t_5$ distribution are found using a maximum-likelihood
technique which avoids binning the data and takes into account the
completeness function and the photometric error as a function of
magnitude.  The quoted mean parameter values and their errors are from
one-dimensional projections of the two-dimensional maximum-likelihood
distribution.  As found by \cite{secker92} and \cite{sh93}, the $t_5$
distribution,
\begin{equation}
\gamma(m) = \frac{8}{3\sqrt{5}\pi\sigma_t}\left[ 1 + 
\frac{(m-m^0)^2}{5\sigma_t^2} \right]^{-3},
\end{equation}
gives a better fit than a Gaussian distribution because the power-law
form of the wings of the $t_5$ distribution matches the GCLF better
than the exponential form of a Gaussian.  Therefore, we only quote
the results of the $t_5$ fits.

\begin{figure}
\epsscale{0.9}
\plotone{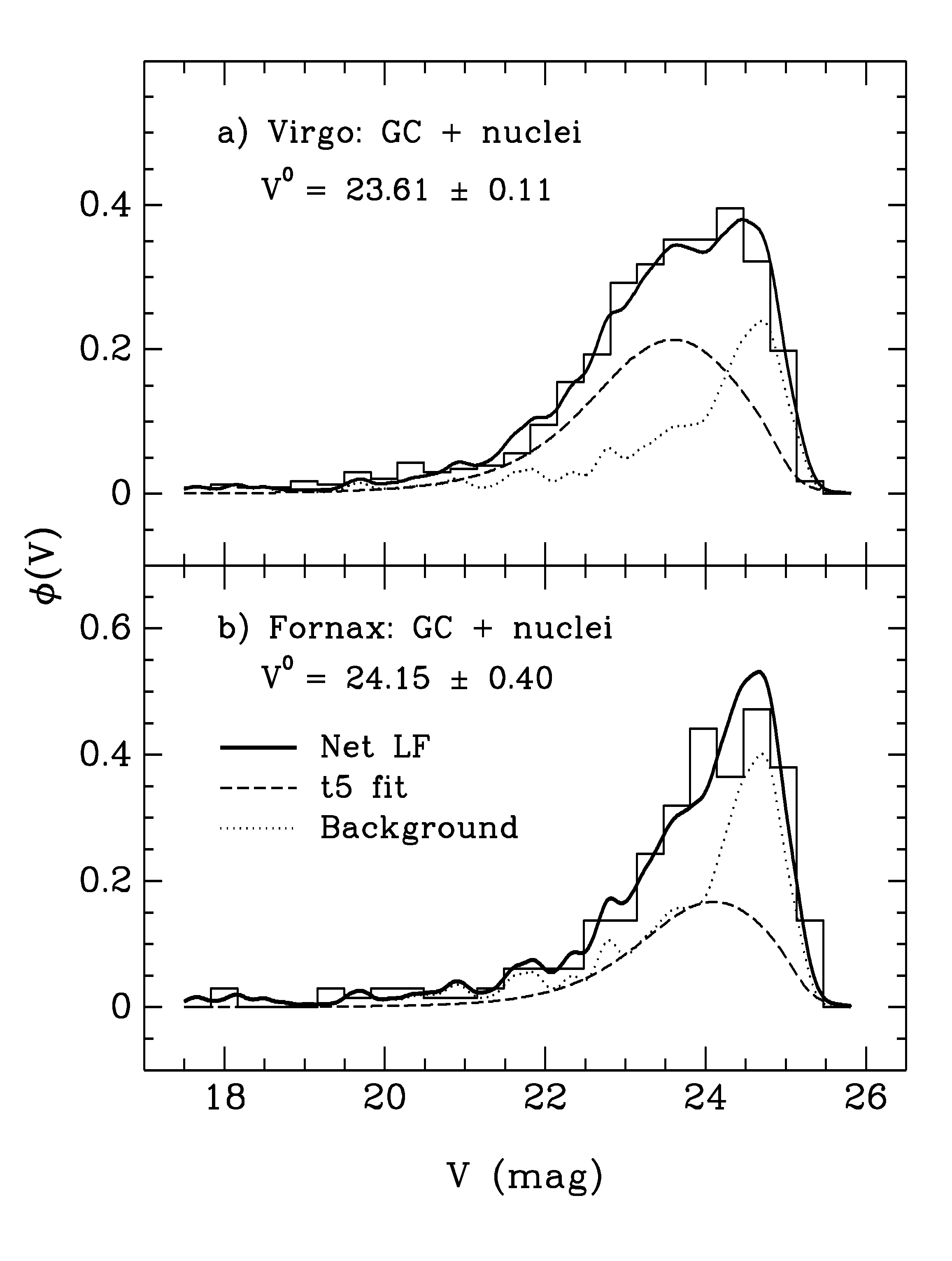}
\caption{$V$-band GC luminosity functions for the Virgo (a) and Fornax
 (b) complete samples (GC plus nuclei). The histograms are the LFs for
 all GC candidates.  The dotted curve is the scaled LF of background
 objects.  These values are given in Tables~\ref{tab:gclfvgo} and
 \ref{tab:gclffnx}. The dashed curve is the best-fitting $t_5$
 distribution using the method of \cite{sh93} and the dark solid curve
 is the total fit.  All the smooth curves have been convolved with the
 completeness functions in Figure~\ref{fig:cplt}.}
\label{fig:gclfallv}
\end{figure}

\begin{figure}
\epsscale{0.9}
\plotone{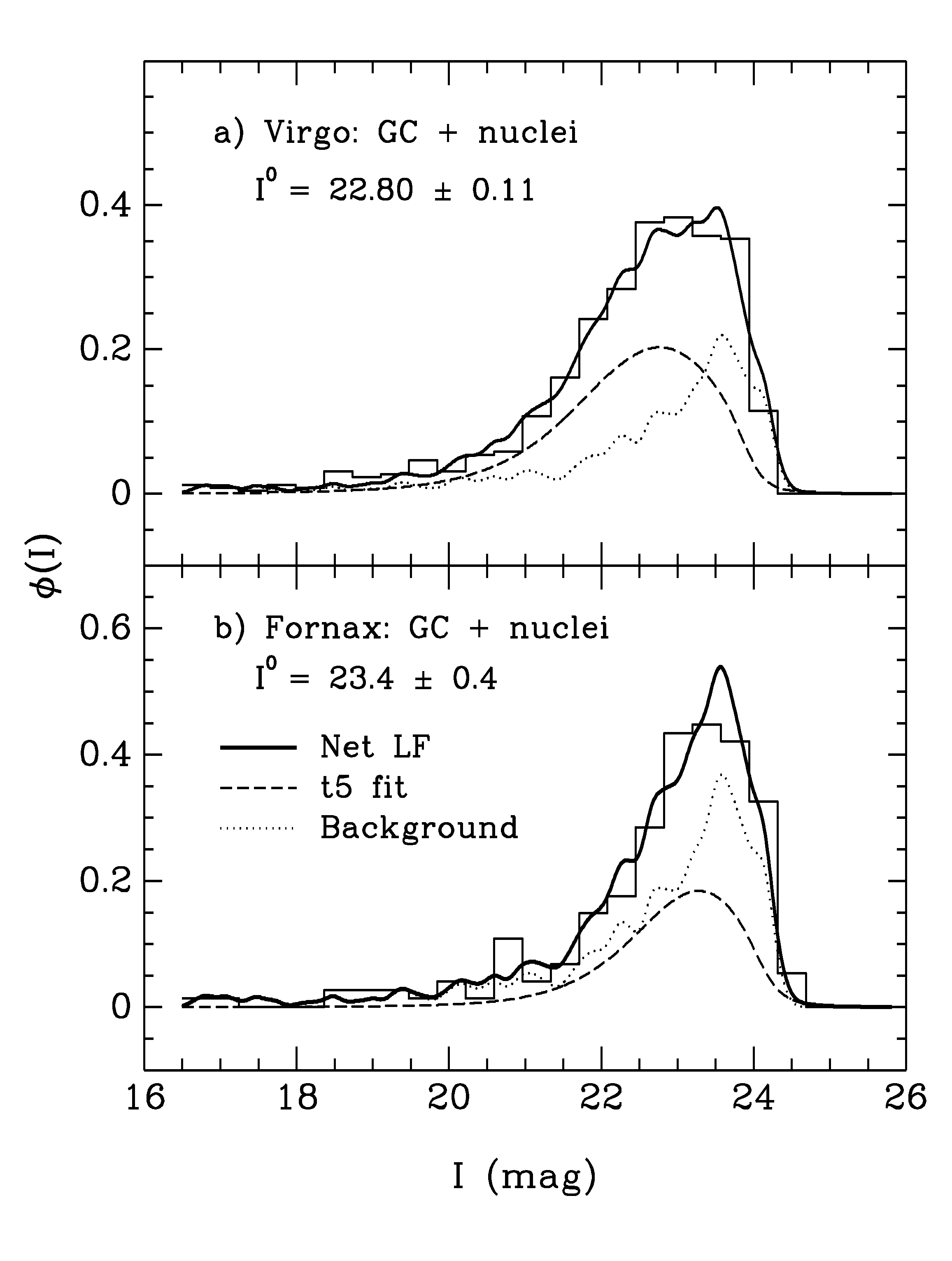}
\caption{$I$-band GC luminosity functions for the Virgo (a) and
 Fornax (b) complete samples (GC plus nuclei). The histograms are the
 LFs for all GC candidates.  The dotted curve is the scaled LF of
 background objects.  These values are given in
 Tables~\ref{tab:gclfi}. The dashed curve is the best-fitting $t_5$
 distribution using the method of \cite{sh93} and the dark solid curve
 is the total fit.  All the smooth curves have been convolved with the
 appropriate completeness functions.}
\label{fig:gclfalli}
\end{figure}

\cite{lmf04} showed that the integrated properties of dE GC candidates
and dE nuclei are very similar.  Photometrically nuclei are like
bright, somewhat red, globular clusters.  Therefore, we first
calculate the GCLF of GC candidates and nuclei together.  The GCLFs
are given in Tables~\ref{tab:gclfvgo}, \ref{tab:gclffnx}, and
\ref{tab:gclfi} and are shown in Figures~\ref{fig:gclfallv} and
\ref{fig:gclfalli}. The parameters of the best fits are given in
Table~\ref{tab:gclf}.  The parameters for the candidates in the Virgo
Cluster are much more accurate than the parameters from the Fornax
Cluster because the Virgo Cluster is 0.5~mag closer than Fornax, so
clusters are detected further down the GCLF, and because the Virgo
sample is larger.  Both the absolute peak magnitudes and the widths
are the same within the errors with mean values of $M_V^0 \approx -7.3
\pm 0.1$ and $M_I^0 \approx -8.1 \pm 0.1$ and $\sigma_{t,} \approx 1.0
\pm 0.1$~mag.  In theory the equivalent Gaussian width is $\sigma_G =
1.29\sigma_t \approx 1.3$~mag.  However, in practice the widths from
Gaussian fits are narrower than the theory would predict. A Gaussian
fit to the Virgo GC$+$nuclei sample gives $m_V^0 = 23.61 \pm 0.13$
with $\sigma_G = 1.16 \pm 0.15$. This is consistent with the $t_5$ fit
and narrower than expected analytically.

\begin{figure}
\includegraphics[width=\textwidth]{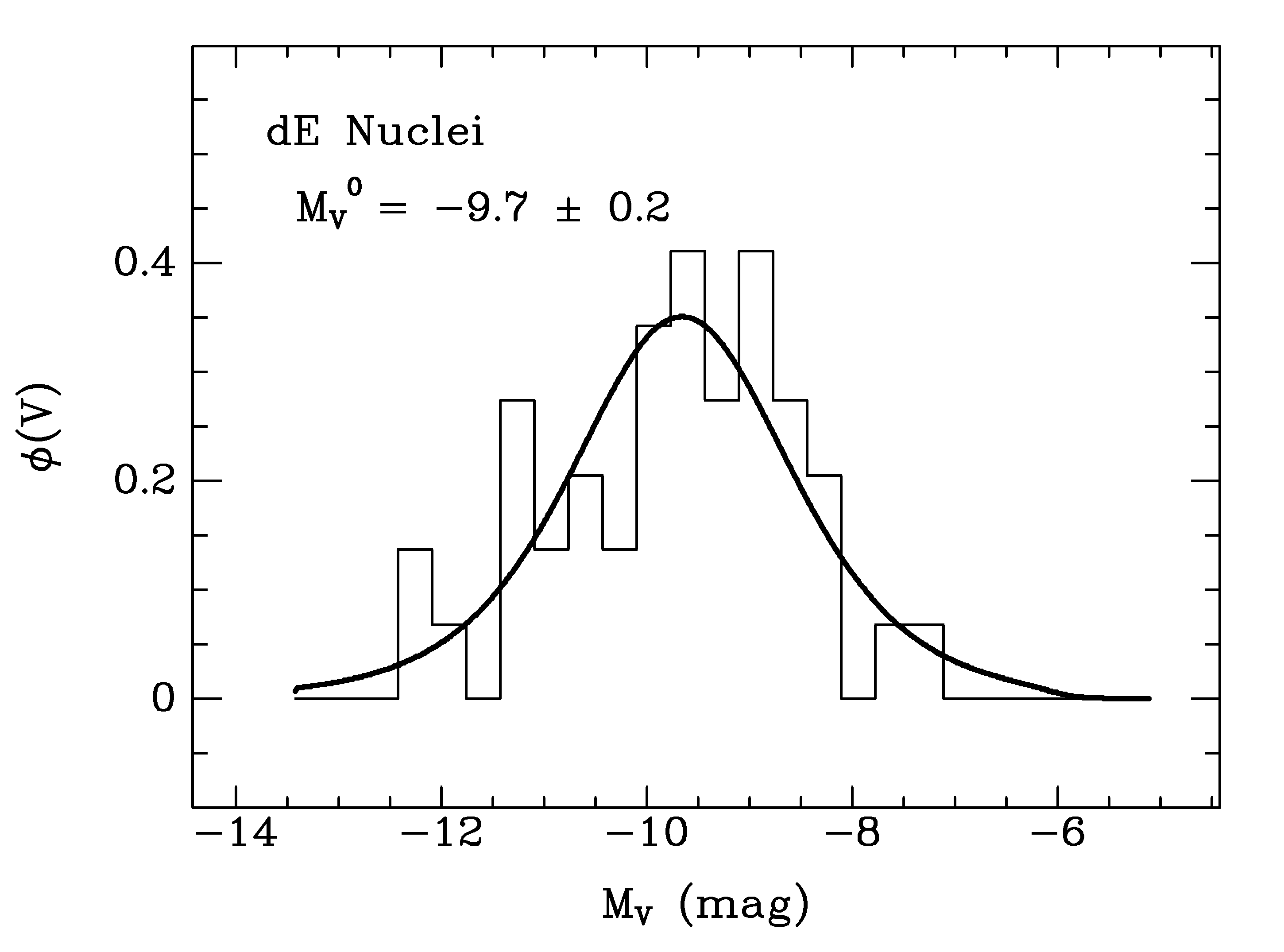}
\caption{$V$-band luminosity function of the dE nuclei.  Since the
  probability that a bright, compact source near the center of a dE
  would be a foreground or background source is small, no background
  subtraction has been done. The solid curve is the best-fitting $t_5$
  function with a peak of $M_V^0 \approx -9.7$
  (Table~\ref{tab:gclf}).}
\label{fig:lfnuc}
\end{figure}

It is also of interest to look at the luminosity functions of the GC
candidates without the nuclei and of the nuclei themselves.  The GCLF
parameters for the GC sample that excludes the nuclei are given in
Table~\ref{tab:gclf} and they are the same, within the errors, to the
sample with GCs plus nuclei.  Figure~\ref{fig:lfnuc} shows the
luminosity function of the nuclei themselves.  For this fit the
magnitudes of the nuclei in the Fornax galaxies have been adjusted to
the distance of the Virgo Cluster.  The LF of the nuclei is peaked
with $M_V^0 \approx -9.7$ (Table~\ref{tab:gclf}).   

In the Virgo Cluster there are a sufficient number of galaxies that we
can also look for differences in the GCLF between nucleated and
non-nucleated galaxies.  The GCLF for dE,N galaxies, excluding the
nuclei, gives $m_V^0 = 23.69\pm0.11$ (Table~\ref{tab:gclf}).
Including the nuclei gives an indistinguishable result of $m_V^0 =
23.66\pm0.13$. Not surprisingly given that the majority of the total
sample of Virgo GCs come from dE,N, the GCLF peak for the dE,N
galaxies of $M_V^0 = -7.23 \pm 0.11$ is similar to the result for the
full sample (Table~\ref{tab:gclf}).  The peak magnitude for the dE,noN
galaxies is $M_V^0 = -7.50 \pm 0.17$, 0.2 magnitudes brighter than
that of the dE,N galaxies, but this difference has a significance of
only 1.2 sigma.  Therefore, $V$-band the GCLFs in dE,Ns and dE,noNs
are the same within the uncertainties.

Thus far we have assumed that all the galaxies in sample are at the
mean distances of their respective groups or clusters.  However, each
of these physical structures has some depth, recent distance
measurements of galaxies in the Virgo Cluster have shown that it has a
line-of-site depth of between 3 and 4~Mpc \citep{jerjen04,jordan05}.
Therefore, we have run Monte-Carlo simulations of composite GCLFs
drawn from 69 galaxies drawn from Gaussian distance distributions with
a mean of 15.3~Mpc and sigma parameters between 1~Mpc and 4~Mpc.  The
distribution of galaxy absolute magnitudes is the same as for the
actual sample.  The number of simulated clusters per galaxy is derived
from a fit to $N_{\rm GC}({\rm tot})$ vs.\ $M_B$, modulated for
Poisson statistics The distribution of simulated GC brightness comes
from a $t_5$ distribution with $M_V^0=-7.3$ and $\sigma_t=1.1$ at our
assumed distance of the Virgo Cluster.  The parameters of the combined
GCLFs were measured with the same procedure used for the real data.
The results of the simulations are shown in Table~\ref{tab:gclfsim}.
As the line-of-sight distribution of galaxies increases the measured
peak becomes systematically brighter and broader than the actual
distribution.  However, as long as the sigma of the galaxy
distribution is less than 2~Mpc (corresponding to a line-of-sight FWHM
of 4.7~Mpc), the systematic errors are not significant.  Also, none of
the galaxies in our sample are members of substructures such as the W,
W$^{\prime}$, or M clouds that are more distant than the main Virgo
Cluster. Therefore, we conclude that our results for $M_V^0$ and
$\sigma_t$ are not significantly biased by the finite depths of the
Virgo and Fornax Clusters.

A log-normal or Gaussian luminosity distribution per unit magnitude
can also be described by a broken power-law when converted to number
per unit luminosity \citep{hp94}.  Figure~\ref{fig:gclf2}a shows the
background-subtracted luminosity function $\log(\phi(L)/L)$ versus log
luminosity for GCs and nuclei in the Virgo sub-sample.  The
best-fitting $t_5$ function from Figure~\ref{fig:gclfallv}\ is plotted
with the dashed line. Above $\log(L/L_{\sun}) \approx 5$ the GCLF is
well-fit with a power-law with slope $\alpha = -1.9\pm0.1$ (straight
black line).  The result is the same if the nuclei are excluded.  This
slope is consistent with the bright-end GCLF slopes for young star
clusters \citep[e.g.][]{whit99} and old GCs in spiral and elliptical
galaxies \citep{hp94}. 

Recently \cite{vdb06} claimed that the break in the power-law GCLF is
present only for galaxies brighter than $M_V = -16$ but that the GCLF
for fainter galaxies is an unbroken power-law. Figure~\ref{fig:gclf2}b
shows the GCLF (GC $+$ nuclei) for Virgo galaxies with $M_V > -15.75$
and the values are given in Table~\ref{tab:gclfvgo_faint}.  The limit
of $M_V = -15.75$ was chosen because there is a natural break in the
distribution of $M_V$ for the WFPC2 sample at this value. The
best-fitting power-law has a slope $\alpha = -1.7 \pm 0.2$ (solid
line), consistent with that for the full sample. Because of the
smaller sample size there is less evidence for a break in the
power-law.  The best-fitting $t_5$ function (dash-dot line) has a
brighter peak and a much broader $\sigma_t$ than the best fit for the
full sample.  However, the standard fit for the full sample (dashed
line) is still a reasonable match to the data.  Deeper data and a
bigger sample are needed to confirm whether the GCLF for faint
galaxies in the Virgo Cluster is different from the GCLF for the
brighter galaxies.

\subsection{Specific Frequency}

Specific globular cluster frequency, \sn, is defined as the number
of globular clusters per unit $V$-band luminosity, normalized at $M_V
= -15$:
\begin{equation}
S_{\rm N} = N_{\rm GC}({\rm tot})10^{0.4(M_V + 15)}
\end{equation}
where $N_{\rm GC}({\rm tot})$ is the total number of globular clusters
\citep{hvdb81}.  With the completeness function and luminosity
function known we can correct $N_{\rm GC}$ for the undetected portion
of the GCLF and then calculate \sn.  The completeness corrections are
determined by dividing the integrals of the measured GCLFs alone by
the integrals of the GCLFs multiplied by the completeness functions.
These factors are 1.04, 1.17, and 1.24 for Leo, Virgo, and Fornax,
respectively.  For the galaxy absolute magnitude we use $M_V = M_B -
0.77$, where 0.77 is the mean $B-V$ color for the dE galaxies in the
study of \cite{vz04}.  The \hst\ data itself cannot be used to
determine the total $V$ magnitude in all cases since the
signal-to-noise for many of the fainter galaxies is not enough to give
a reliable magnitude.

The computed values for $N_{\rm GC}({\rm tot})$ and \sn\ are given in
Tables~\ref{tab:denprop} and \ref{tab:denonprop} and plotted in
Figure~\ref{fig:snmv}.  The overall trends noticed by \cite{mil98} in
a subset of the current data are still apparent, \sn\ increases with
increasing $M_V$ (decreasing luminosity).  Also, on average nucleated
galaxies have higher \sn\ than non-nucleated galaxies: $\bar{S}_{\rm
N}({\rm dE,N}) = 8.7\pm1.6$ and $\bar{S}_{\rm N}({\rm dE,noN}) =
4.3\pm1.1$ where the uncertainties are the standard deviations of the
mean.  These values are slightly higher than the values in
\cite{mil98} since the current, larger sample contains more faint
galaxies which typically have higher values of \sn.

Since \sn\ is correlated with galaxy absolute magnitude, it is also
correlation with any other property that depends on the galaxy
brightness.  For example, it is well known that the surface brightness
profiles of faint dEs have Sersic exponents near 1 (exponential) while
brighter dEs are more cuspy with exponents approaching 4, equivalent
to an $R^{1/4}$ profile \citep{gg03}.  Put another way, more luminous
dEs have higher surface brightnesses. Figure~\ref{fig:snsb} shows how
the scatter in \sn\ increases with decreasing central surface
brightness for the sub-sample of the current sample in
\cite{stiavelli01}.  The mean central surface brightness from Sersic
fits for 14 dE,N galaxies is $\mu_0(V) = 21.3 \pm 1.6$ while for 10
dE,noN galaxies it is about 1 magnitude fainter, $\mu_0(V) = 22.2 \pm
1.5$.

There is also an environmental dependence on \sn\ in the sense that
the number density of dE,N galaxies is more centrally concentrated
than that of the bright dE,noN galaxies in the Virgo and Fornax
Clusters \citep{fs89}.  However,  there is no direct radial dependence
on \sn\ with projected radial position within the Virgo Cluster or with
projected radial distance from the nearest bright ($M_B < -18.6$)
galaxy (Figure~\ref{fig:vgoradsn}).

\subsection{Globular Cluster T Parameter and Mass Fractions}

\sn\ is a very useful quantity since it is dependent only on basic
measurements and it has a long history in the literature for
comparison.  However, because it is based on a luminosity it is
applicable only to old stellar systems with similar mass-to-light ($M/L$)
ratios.  It is important to be able to compare the star cluster
systems in different types of galaxies, so a quantity like \sn\ is
needed that is independent of $M/L$.  Such a parameter is the $T$
parameter of \cite{za93}, 
\begin{equation}
T=\frac{N_{\rm GC}({\rm tot})}{M_G/10^9 M_{\sun}}
\end{equation}
where $N_{\rm GC}({\rm tot})$ is the total number of clusters and
$M_G$ is the stellar mass of the galaxy.  $M/L$ values are not
available for each galaxy so we have adopted $M/L_V = 5$, a typical
value for dE galaxies from the spectroscopic measurements of
\cite{geha02}.  Values of \tv\ for the current sample are given in
Tables~\ref{tab:denprop} and \ref{tab:denonprop}.  Not surprisingly,
Figure~\ref{fig:tvmgal} shows that \tv\ exhibits the same trends with
galaxy mass as \sn\ has with $M_V$.  Likewise, the mean \tv\ for the
nucleated galaxies is about a factor of two higher than the mean for
the non-nucleated galaxies: $\bar{T}({\rm dE,N}) = 20.4\pm3.9$ and
$\bar{T}({\rm dE,noN}) = 10.2\pm2.6$. 

Finally, it is of interest to estimate the mass fraction of the
stellar light currently in star clusters.  Because of the
uncertainties in estimating the total mass in GCs in any galaxy given
the small number statistics, we calculate the mass in two ways and
average the results.  In both cases we assume $M/L_V = 2$ for each GC.
The ``specific mass'', or percentage of stellar mass in GCs is
\begin{equation}
F = 100\frac{M_{GC}}{M_G}. 
\end{equation}

In the first method we take advantage of the fact that we believe that
we know the form of the mass function.  The total mass of clusters
should then be roughly equal to the mean cluster mass times the number
of clusters.  Integrating over $M\phi(M)$, $\phi(M)=\phi(L)*M/L_V$
where $\phi(L)$ is the $t_5$ function shown in Figure~\ref{fig:gclf2}a
and we assume $M/L_V = 2$, gives a mean mass of
$M=4.3\times10^5$~\Msun.  The mass fraction then is given by $F =
0.0433T$.

In the second method, the magnitude of each object in the
``object+background'' sample is converted to a mass assuming $M/L_V =
2$.  Second, from the total (GC + background) and background
luminosity functions we compute a probability that an object of a
given magnitude is an actual GC rather than a background object,
\begin{equation}
P(V)=1-\frac{\phi_{bkg}(V)}{\phi_{tot}(V)}.
\end{equation}
For example, in Figure~\ref{fig:gclfallv} the $\phi_{tot}(V)$ are the
thick solid curves (normalized column 2 of Tables~\ref{tab:gclfvgo}
and \ref{tab:gclffnx}) and the $\phi_{bkg}(V)$ are the dotted curves
(column 3 of Tables~\ref{tab:gclfvgo} and \ref{tab:gclffnx}). A fourth
order polynomial is fit to $P(V)$ in order to smooth it and the
results for the Virgo and Fornax samples are shown in
Figure~\ref{fig:gcprob}.  The total mass of clusters for each galaxy
is then the sum of the cluster masses times the probability that each
is an actual cluster.  No additional correction is made for the
unobserved portions of the GCLFs, but this should be a small factor
since most of the mass is in the most massive clusters.

The two methods give results that are consistent within about a factor
of two.  The mean values of \tm\ from these two methods and their
uncertainties are also in Tables~\ref{tab:denprop} and
\ref{tab:denonprop}.  The trends in \tm\ with galaxy mass are similar
to those seen with the other parameters (see Figure~\ref{fig:tmmgal})
and the mean values are $\bar{F}({\rm dE,N}) = 0.9\pm0.2$ and
$\bar{F}({\rm dE,noN}) = 0.7\pm0.3$.  The mean for the non-nucleated
dwarfs is biased by one outlying point, VCC~1781 with $F = 7.5\pm7.8$.
Removing this point gives $\bar{F}({\rm dE,noN}) = 0.5\pm0.1$.

\section{Discussion}\label{sec:disc}

\subsection{Globular Cluster Luminosity Functions}

An important question is whether the globular cluster luminosity and
mass functions in dwarf galaxies are the same as those in spiral or
giant ellipticals.  The comparison will have implications for theories
of globular cluster formation and evolution and will put constraints
on the fraction of globular clusters in different types of giant
galaxies that were formed in dwarf galaxies and later accreted.  First
we compare our results on the form of the GCLF and GC mass function in
dEs with previous authors and then address the issue of how the GCLF
of dEs compares with other types of galaxies.

Our best measurement of the shape of the GCLF comes from the Virgo
sample. The $V$-band GCLF for the Virgo dEs is well-fit by a $t_5$ or
Gaussian function with $M_V^0=-7.3\pm0.1$ and $\sigma_t = 1.0 \pm 0.1$
mag (as mentioned above, a Gaussian fit give $\sigma_G = 1.2 \pm
0.2$).  \cite{durrell96} measured the GCLF peak from a sample of 11
Virgo dEs to be $T1^0 = 23.6\pm0.3$.  With $(V-T1) = 0.5$ and using
our adopted distance modulus of 30.92 for Virgo, this gives
$M_V^0=-6.8\pm0.3$ which is not consistent with our results.
\cite{sharina05} have recently studied the GC systems of a sample of
dwarfs with distances between 2 and 6~Mpc with \hst.  The GCLF for the
dEs has a peak at $M_V^0 \approx -7.4$ but at $M_V > -6$ the numbers
continue to rise instead of decrease.  Our data do not go deep enough
to investigate the shape of the faint end of the GCLF. Also,
\cite{beasley06} have studied the system of $\sim77$ GCs in the Virgo
dE VCC~1087 and found that $g^0 = 24.1 \pm 0.3$.  With $V \approx
g-0.3$ \citep{bc03}, this gives $M_V^0 = -7.1$ for $(m-M)_0 = 30.92$
(mean Virgo distance modulus) or $M_V^0 = -7.5$ for $(m-M)_0 = 31.27
\pm 0.14$ \citep[SBF distance]{jerjen04}.  Therefore, this result is
consistent with ours within the uncertainties in the peak measurement
and the distance.

In the $I$-band we find $M_I^0 = -8.1 \pm 0.1$ for the Virgo Cluster
dE ``GC $+$ nuclei'' GCLF and $M_I^0 = -8.0 \pm 0.4$ for the Fornax
Cluster dE ``GC $+$ nuclei'' GCLF.  These values are fainter than
the GCLF turnover luminosity for giant ellipticals, $M_I^0 =
-8.46 \pm 0.03$, found by \cite{kundu01a}. 

The literature to date has been somewhat undecided about whether the
GCLF of dE differs from that of giant ellipticals.  Figure~2 of the
review of \citep{harris91}, which includes only the four Local Group
dEs with GCs known at the time, shows a small but statistically
insignificant trend that the GCLF peak in dwarfs may be fainter than
in giants.  Based on the statistics available Harris concluded that
the GCLF peak is fairly independent of galaxy mass.  \cite{durrell96}
found that the the peak from their dE sample was 0.4~mag fainter than
the GCLF peak of M87 from \cite{whit95}.  Note that Durrell's GCLF
peak is about 0.5~mag fainter than is currently
found. \cite{strader06} claim that a comparison of GCLF peaks in dEs
and gEs in Virgo are consistent.  However, \cite{jordan06,jordan07},
using the same ACS Virgo Survey dataset, find that the GCLF peak
magnitudes tend to scatter towards fainter values in lower-luminosity
galaxies.

Recently \cite{dicris06} have done a systematic analysis of the
literature to determine if there are differences in the GCLF peaks
between the GC systems of the Galaxy, M31, and giant ellipticals.  All
distances were put on a consistent scale calibrated to RR~Lyraes and
Cepheids assuming that $\mu_0(LMC) = 18.50$.  Similar selection
criteria are used for the Milky Way and M31 samples. Metallicity
corrections to the Cepheid distances to the galaxies used to calibrate
the surface brightness fluctuation (SBF) method are from theoretical
models rather than the empirical relation used by the \hst\ Key
Project.  This gives better agreement between the Cepheid and SBF
distances.  Finally, metallicity corrections from \cite{acz95} are
applied to the GCLF peak magnitudes so that all peaks are for $[{\rm
Fe/H}] = -1.6$.  The result is that the GCLF peaks the metal-poor GC
populations are very constant with $M_{V,Z}^0(MW) = -7.66 \pm 0.11$,
$M_{V,Z}^0(M31) = -7.65 \pm 0.19$, and $M_{V,Z}^0(gE) = -7.67 \pm
0.23$.  For comparison with the current work, using the Cepheid
calibration from \cite{freedman01} gives $M_{V,Z}^0(gE) = -7.83 \pm
0.23$.

From \cite{lmf04} the mean GC color is $\langle V-I\rangle = 0.90$.
According to the following color-metallicity relation from
\cite{kpatig98}
\begin{equation}
[{\rm Fe/H}] = -4.5 + 3.27(V-I)
\end{equation}
the mean metallicity of the dE GCs is $[{\rm Fe/H}] = -1.6$.  This is
the same as the metallicity of the metal-poor GC samples from
\cite{dicris06}, so no additional metallicity correction to the dE
GCLF peak is needed and $M_{V,Z}^0(dE,HST) = -7.3 \pm 0.1$

A comparison of the GCLF peak results is shown in
Figure~\ref{fig:gclfpeak}.  Except for the three filled circles all of
the values have been adjusted, where appropriate, to use the HST Key
Project distance moduli used in this paper.  The three filled circles
use Di~Criscienzo et al.'s preferred distance scale which implies a
Virgo distance modulus of $(m-M)_0 = 30.76$. In general, the peak
GCLF magnitudes for the dEs are fainter than the peak magnitudes for
the MW, M31, and gEs, though the statistical significance in each
individual result is only 1--1.5 sigma.  The indication from all the
results together is that the GCLF peak in dEs is similar to or fainter
by $\sim0.5$ magnitudes than the peak in giant galaxies.  That our
$I$-band peak magnitudes are also $\sim0.4$ mag fainter than those
measured in giant ellipticals adds support to this result.

\begin{figure}
\plotone{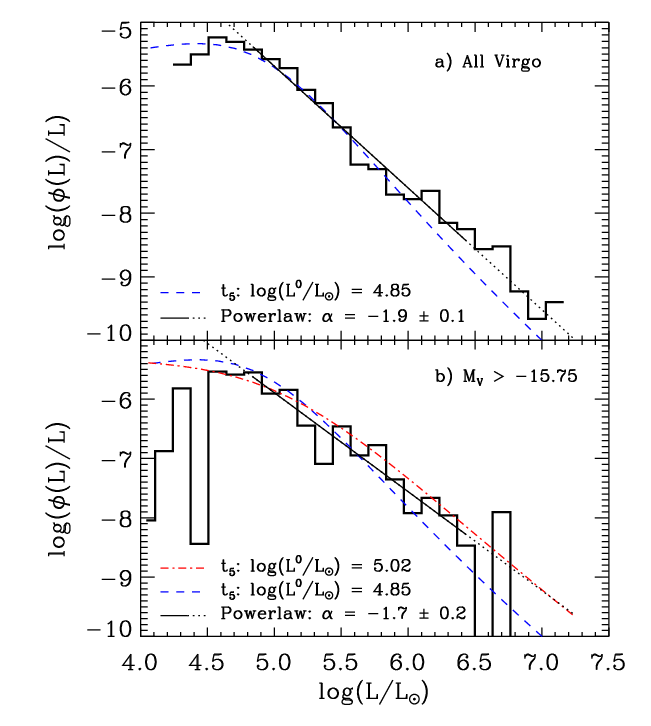}
\caption{Power-law representations of the background-subtracted GCLFs
  for the GC candidates plus nuclei for the Virgo Cluster
  sub-sample. a) All galaxies. b) For faint galaxies with $M_V >
  -15.75$ (see Table~\ref{tab:gclfvgo_faint}). The best-fitting $t_5$
  and power-law fits are shown as straight black lines. The power-law
  fits are made in the interval $4.8 < log(L/L_{\odot}) < 6.5$ show
  with the solid straight lines.  The dotted straight lines show the
  extrapolations of the fits. The dashed line is the best-fitting
  $t_5$ function to the complete Virgo sample (a). The dash-dot line
  is the best-fitting $t_5$ function to the clusters in galaxies with
  $M_V > -15.75$.}
\label{fig:gclf2}
\end{figure}

\begin{figure}
\includegraphics[width=\textwidth]{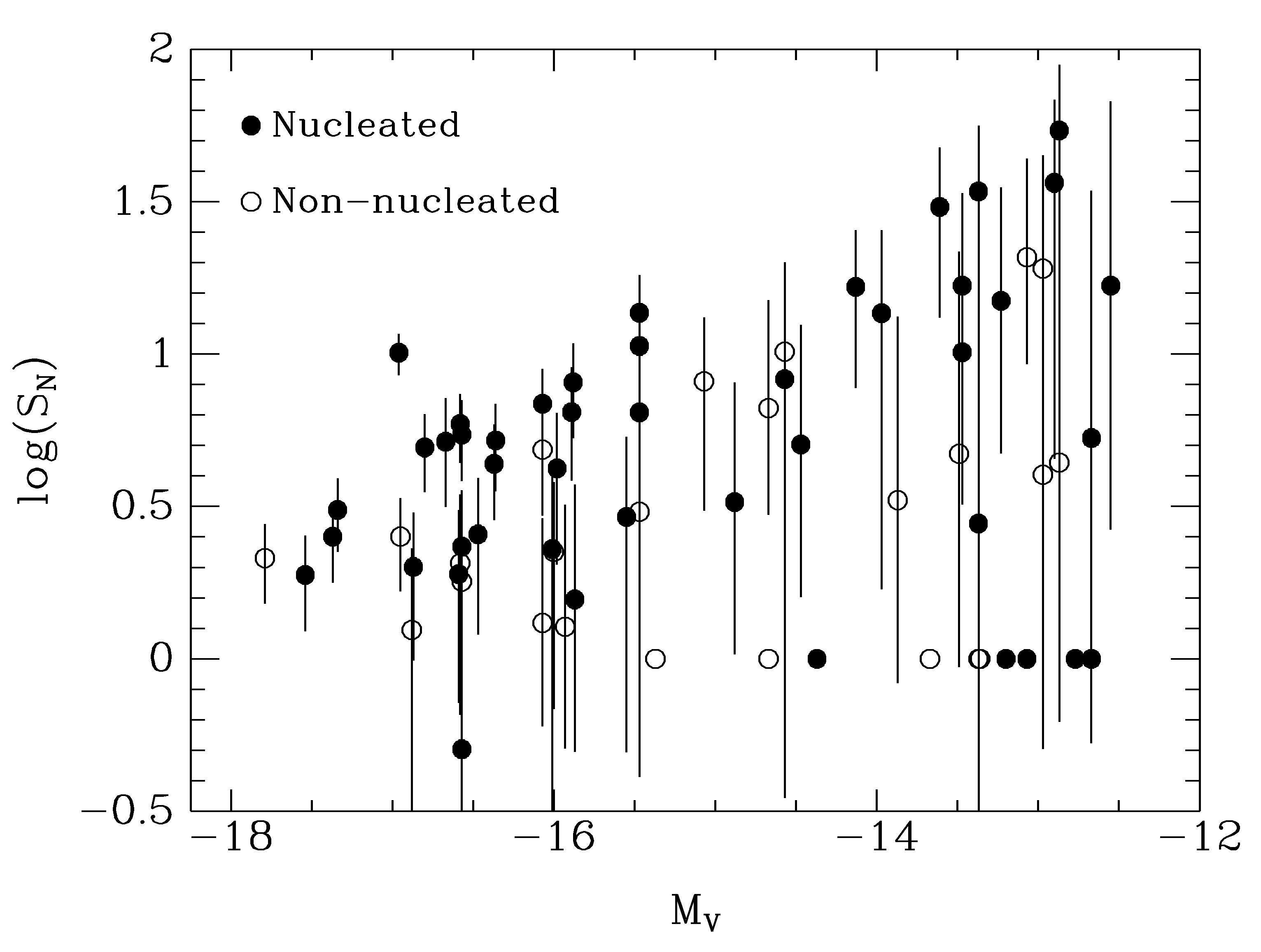}
\caption{\sn\ vs. $M_V$ for the dE,N and dE,noN from the \hst\ dE
  Snapshot Survey.  As seen by \cite{mil98}, \sn\ increases with
  increasing $M_V$ (decreasing galaxy luminosity).  Galaxies with
  $N_{\rm GC} = 0$ have been given $\log(S_{\rm N}) = 0$ so that they
  appear on the plot.}
\label{fig:snmv}
\end{figure}

\begin{figure}
\includegraphics[width=\textwidth]{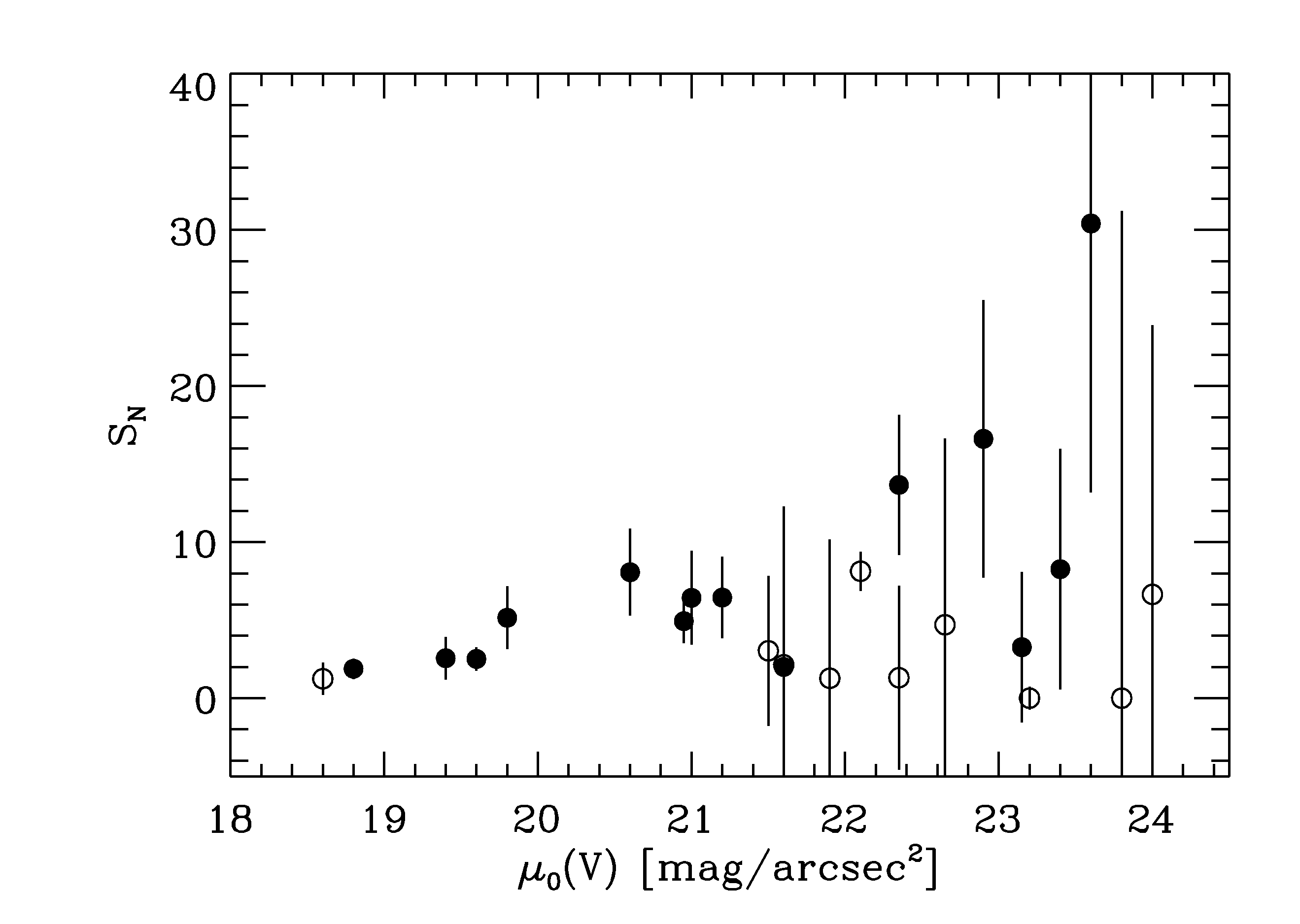}
\caption{\sn\ vs. central surface brightness for dE,N and dE,noN
  galaxies. The central surface brightnesses are from the Sersic fits
  in \cite{stiavelli01} in the radial range from 1\arcsec\ to 5\arcsec\
  and so do not include the nuclei. For the nucleated galaxies \sn\
  increases with decreasing central surface brightness.}
\label{fig:snsb}
\end{figure}

\begin{figure}
\includegraphics[width=0.8\textwidth]{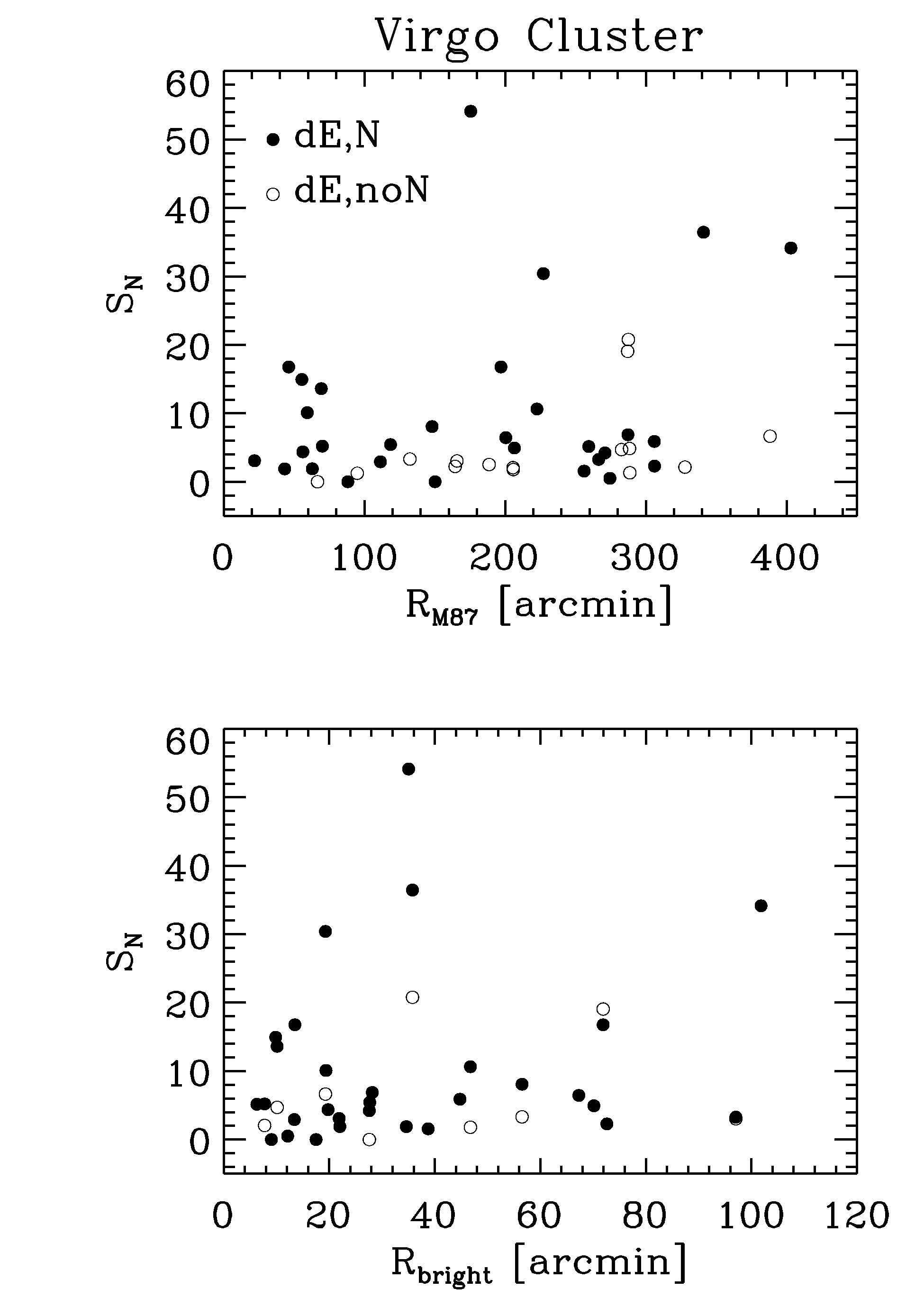}
\caption{\sn\ versus projected radial distance from M87 (upper panel)
  and the projected radial distance from the nearest Virgo galaxy with
  $M_B < -18.6$.  There are no clear trends present but in general
  dE,N always have higher \sn\ than dE,noN at a given radius.}
\label{fig:vgoradsn}
\end{figure}

\begin{figure}
\includegraphics[width=\textwidth]{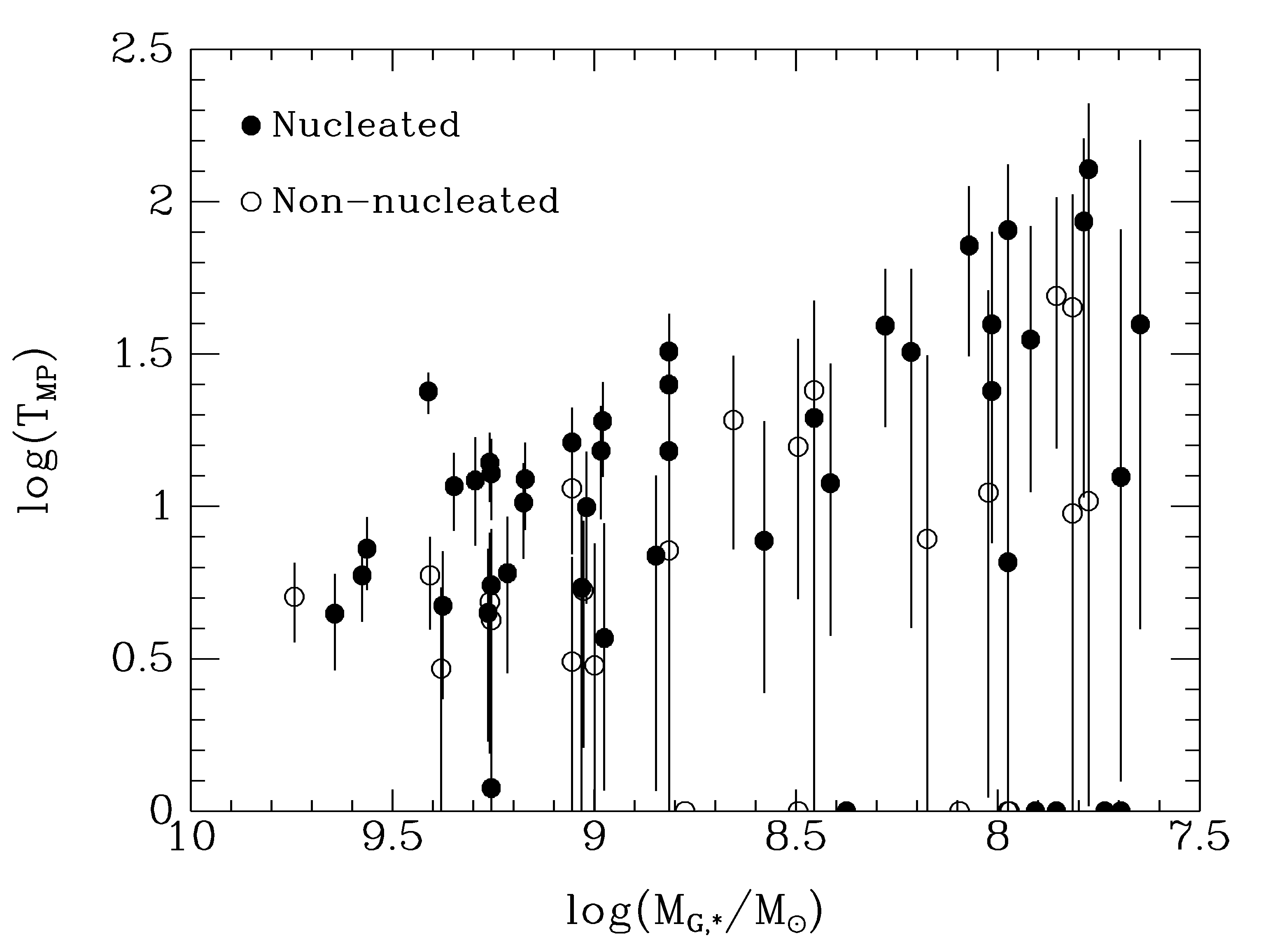}
\caption{\tmp\ vs. $M_{G,*}$ for the dE,N and dE,noN from the \hst\ dE
  Snapshot Survey.  As in Figure~\ref{fig:snmv}, galaxies with $N_{\rm
  GC} = 0$ have been given $\log(T_{\rm MP}) = 0$ so that they appear
  on the plot.}
\label{fig:tvmgal}
\end{figure}

\begin{figure}
\includegraphics[width=\textwidth]{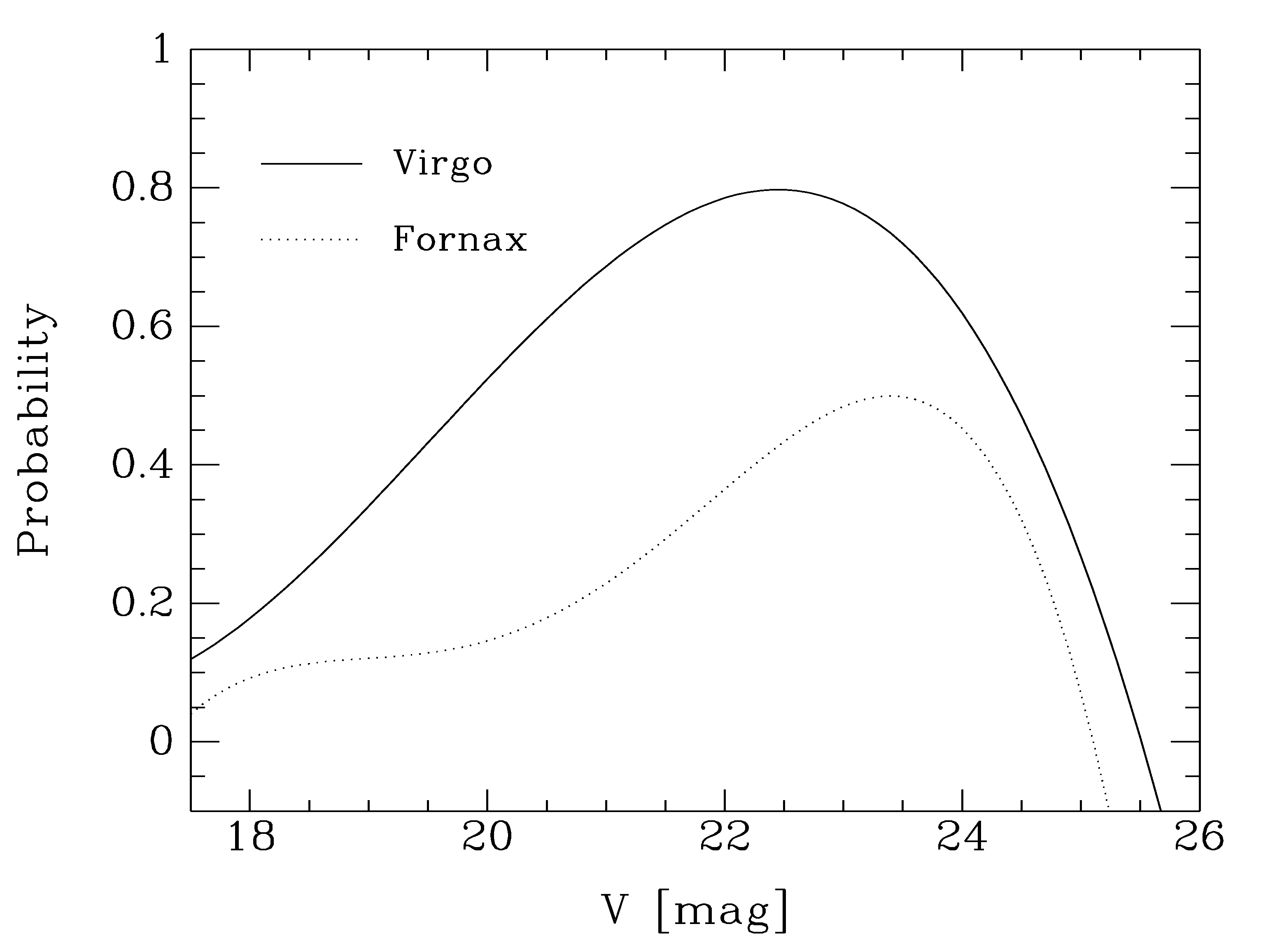}
\caption{Probability with magnitude that a GC candidate in the Virgo
  Cluster sample of a given apparent magnitude is an actual GC, not a
  background object.}
\label{fig:gcprob}
\end{figure}

\begin{figure}
\includegraphics[width=\textwidth]{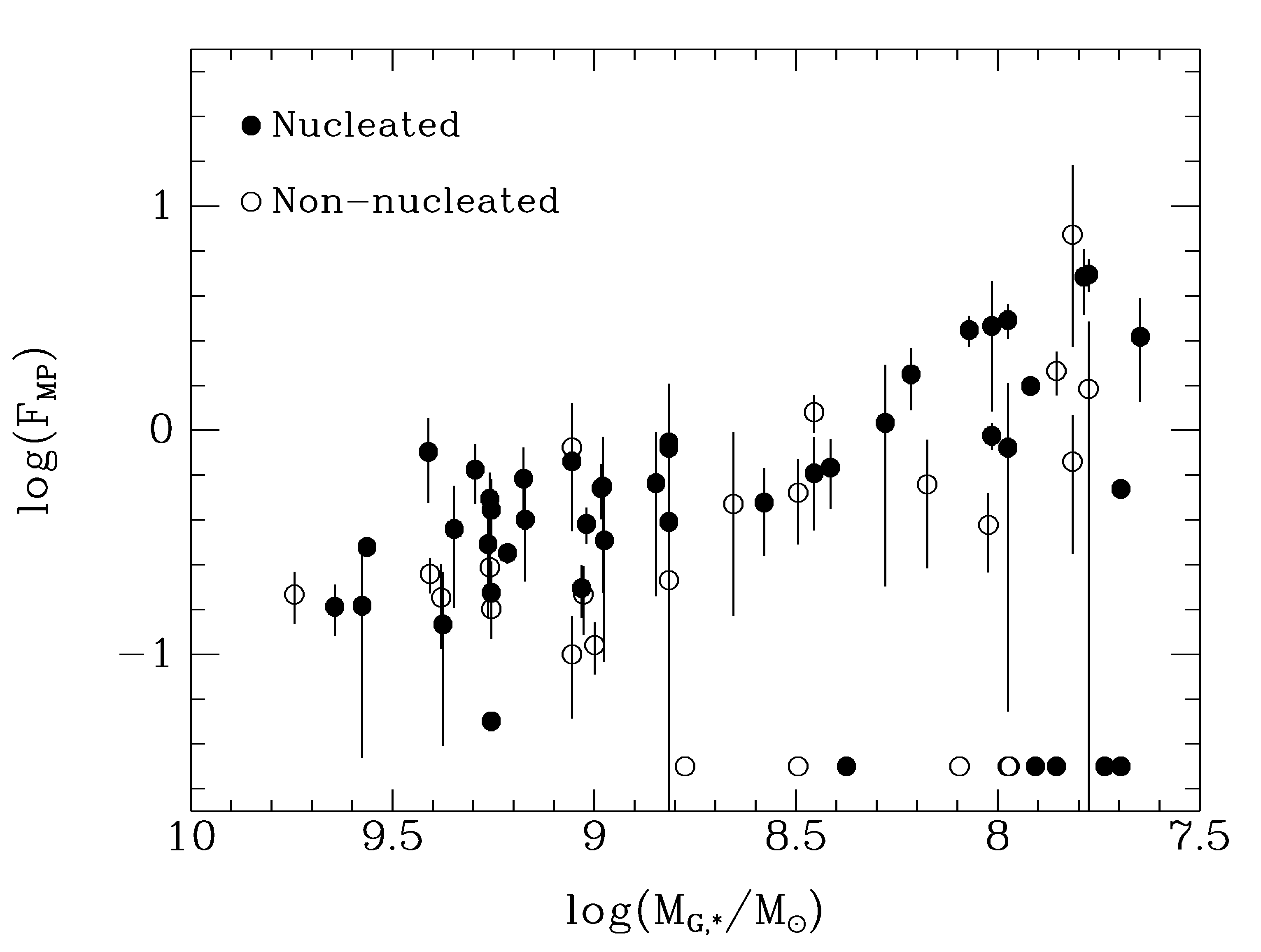}
\caption{\fmp\ vs. $M_{G,*}$ for the dE,N and dE,noN from the \hst\ dE
  Snapshot Survey. Galaxies with $N_{\rm GC} = 0$ have been given
  $\log(F_{\rm MP}) = -1.5$ so that they appear on the plot.}
\label{fig:tmmgal}
\end{figure}

\begin{figure}
\includegraphics[width=\textwidth]{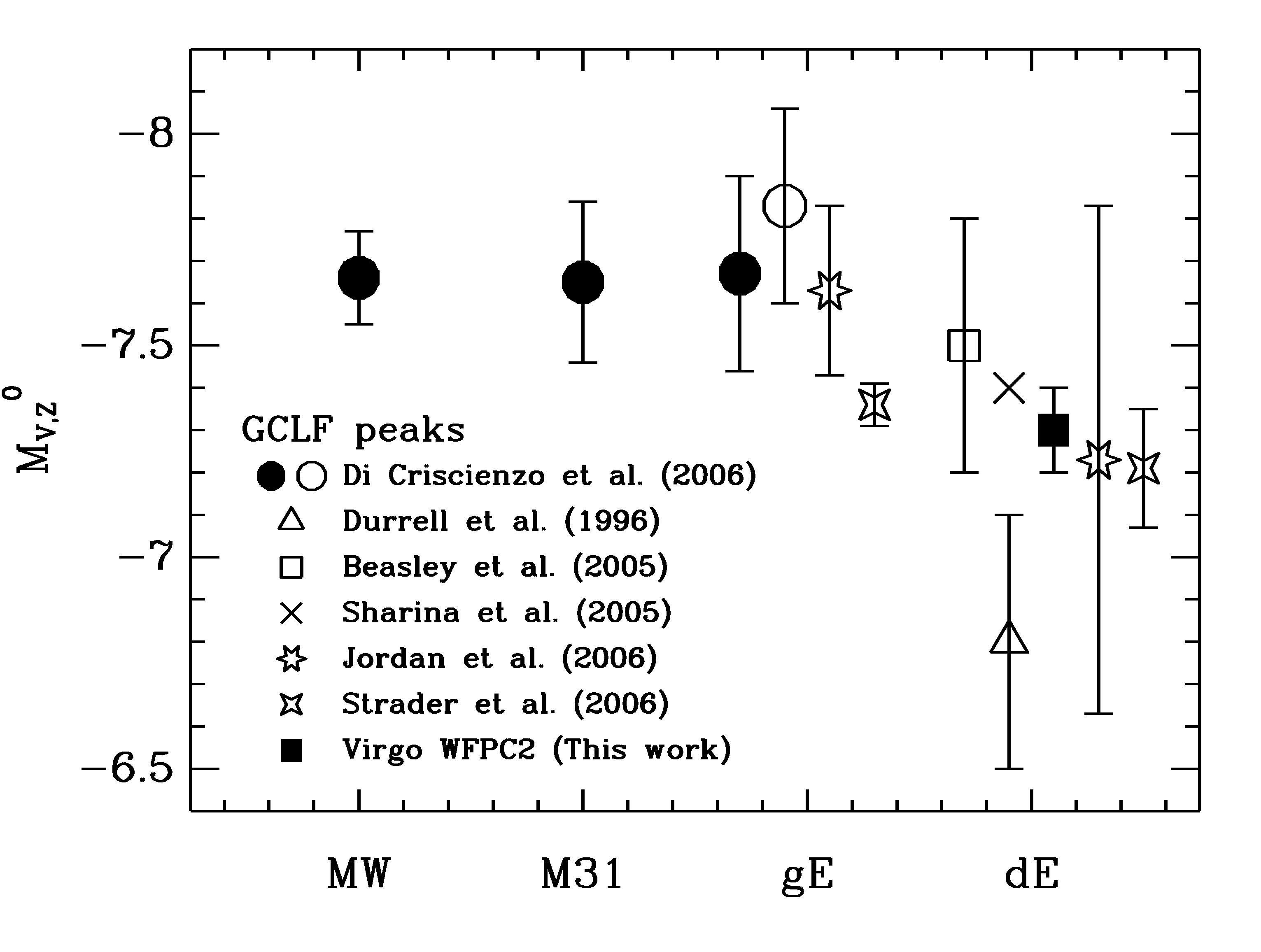}
\caption{Metallicity-corrected GCLF peak magnitudes as a function of
  host galaxy type. The filled circles are the preferred values from
  \cite{dicris06} while the open circle represents the GCLF peak for
  gEs using the Key Project distances \citep{freedman01}. The filled
  square is from the Virgo Cluster sample in this paper and the other
  points are from the literature
  \citep{durrell96,sharina05,beasley06,jordan06,strader06} corrected
  to the Key Project distances where appropriate.  While there is
  significant overlap between the GCLF peak magnitudes in giant and
  dwarf galaxies, the impression is that on average the GCLF peak in
  dEs is 0.2--0.5 magnitudes fainter than in giant ellipticals.}
\label{fig:gclfpeak}
\end{figure}

The shapes and peaks of the observed GCLFs in different types of
galaxies are important for what they reveal about the underlying GC
mass function (GCMF).  As shown in Figure~\ref{fig:gclf2}, peaks in
the GCLF as plotted as number per unit magnitude become bends or
flattenings when plotted as number per unit luminosity or
mass. Assuming $M/L = 2$, then $M_{V,Z}^0(dE,HST) = -7.3 \pm 0.1$
corresponds to $\log(m^0(dE,HST)) = 5.15\pm0.04$ while $M_{V,Z}^0(gE)
= -7.66 \pm 0.11$ gives $\log(m^0(gE)) = 5.29\pm0.05$.  The bright-end
slope of these luminosity or mass functions is consistent with a
power-law $N(L) \propto L^{\alpha}dL$.  In the Milky Way, M31, and gEs
$\alpha$ is typically between $-1.6$ and $-2$
\cite[e.g.][]{hp94,mclaughlin94,larsen01}.  Very young star clusters
in merging and starburst galaxies also have nearly pure power-law
luminosity functions with $\alpha$ between $-1.8$ and $-2$
\cite[e.g.][]{mil97,whit99,goud04}. The bright-end slope of the GCLF
for the entire Virgo dE sample (Figure~\ref{fig:gclf2}a) is
$-1.9\pm0.1$, consistent with with previous results for dEs
\citep{durrell96} and with the bright end slopes in the Milky Way and
M31 and young cluster systems. This slope is somewhat steeper than the
bright-end slopes found in many gEs \citep[see][]{hp94}.

Some interesting results from the Virgo ACS survey are that the peak
GCLF magnitudes, or power-law break masses, and the width of the
Gaussian distribution of magnitudes decrease with decreasing galaxy
luminosity \citep{jordan07}.  As shown in Figure~\ref{fig:gclfpeak},
the GCLF peak magnitude measured from the WFPC2 data agrees with the
ACS value once the two results are put on the same distance scale and
corrected for bandpass.  The equivalent Gaussian width from the WFPC2
data is $\sigma_G=1.16\pm0.15$.  For galaxies with $-18 < M_B < -16$
the ACS result is $\sigma_G\approx0.9\pm0.1$.  The WFPC2 GCLF width is
systematically higher than the ACS result at the $1.5\sigma$
level. This is also reflected in the fact that the bright-end
power-law slope of $\alpha = -1.9\pm0.1$ is consistent with but on
average shallower than the slopes measured for dwarf galaxies in the
ACS sample.

The WFPC2 survey contains galaxies fainter than in the ACS Virgo
survey so we can explore if the trends in peak magnitude and Gaussian
sigma continue.  The formal $t_5$ fit to the faint WFPC2 galaxy subsample
gives a much brighter and broader GCLF than for the bright
subsample. However, as can be seen in Figure~\ref{fig:gclf2}b, this is
not a reliable result.  Due to the small number of GCs the GCLF peak for
the fainter galaxies is consistent with the peak for the  brighter
galaxies.  The width of the distribution for the faint galaxies is
broader, if anything, than the width for the bright galaxies and the
power-law slope is shallower.  This is opposite to the trend suggested
by the ACS results.  Unfortunately, the number of GCs in the faint
galaxies is too low to determine if this difference is significant. 

Theoretically it is also not certain whether a trend of GCLF peak, or
mass function turnover, with host galaxy mass is expected.  The
similarities between the bright end of the GCLF for old, metal-poor
clusters and the power-law mass functions of young star clusters and
giant molecular clouds has led to the idea that star clusters form
with an initial power-law mass function which evolves with time as
low-mass clusters are destroyed \citep{hp94,go97}.  \cite{fz01} show
that if $M_{gal} \propto V_c^3$ then the mass function is independent
of $V_c$ while if $M_{gal} \propto V_c^4$ then the peak mass becomes
{\it larger} for smaller $V_c$.  However, \cite{vz03} find that the
mean cluster mass of evolved populations does increase with host
galaxy mass in rough agreement with the peak masses given above if the
cutoff in the stellar IMF is $m_T = 0.9$~\Msun.  One difference
between the two works is that \cite{vz03} give their simulated
clusters King model concentration indices that scale with cluster
mass.  However, all of the model host galaxies have $\log(M_{gal}) >
10.5$ and so are not true dwarfs.  Earlier simulations with dwarf
galaxy-sized objects predict mean cluster masses smaller than given by
the current observations \citep{vesp00,vesp01}.  New simulations for
dwarf-galaxy size hosts would be very useful.

\subsection{Specific Frequencies and Mass Fractions}

The high average \sn\, the trend of increasing \sn\ with decreasing
galaxy luminosity, and the factor-of-two difference in \sn\ between
nucleated and non-nucleated dwarfs all confirm the results of
\cite{mil98} using the first 24 galaxies observed for the \hst\ dE
Snapshot Survey.  Further, a Kolmogorov-Smirnov test comparing the
distributions of \sn\ in nucleated and non-nucleated galaxies gives
that the probability that they are drawn from the same distribution is
only 3\% (see below).  There do seem to be systematically more GCs per
unit luminosity in dE,N than in dE,noN galaxies not accounted for by
the presence of nuclear star clusters alone.

One evolutionary pathway for dEs is the stripping of gas from dIs as
they encounter the dense intercluster medium in rich galaxy clusters.
An important way of evaluating the importance of this mechanism is to
compare the GC populations of dIs and dEs.  New studies of the star
clusters in dIs now allow us to make comparisons using roughly equal
numbers of galaxies \citep{seth04,sharina05}. However, because dIs and
dEs have different $M/L$ ratios, the luminosity-based \sn\ is not the
appropriate quantity to use for the comparison.  Instead, we correct
for $M/L$ by calculating the $T$ parameter for metal-poor GC
populations, $T_{\rm MP}$, to compare the relative numbers of GCs in
dIs, dEs, spirals and giant ellipticals.  In this case we are only
comparing the number of old (age $\gtrsim 10$~Gyr), metal-poor ($[{\rm
Fe/H}] \lesssim -1$) clusters.  We are assuming that all dE GCs are
old and metal-poor. From the sample of Fornax and Virgo Cluster dIs
from \cite{seth04} we use $N_{red}$, the number of cluster candidates
with $V-I > 0.85$.  Background subtraction and completeness
corrections are done in the same way as described in that paper.  The
galaxies' absolute magnitudes are from their Table~1.  We also include
the 11 Virgo dEs and NGC~3115D1 whose clusters were identified by
\cite{durrell96} and VCC~1087 \citep{beasley06}.  When calculating the
luminosities of these galaxies we use the distance moduli assumed
throughout this paper. \cite{sharina05} identified GCs in \hst\
snapshots of 57 dwarf galaxies with distances between 2 and
6~Mpc. From their Table~2 we select only those star cluster candidates
with $0.7 < V-I < 1.5$ in order to exclude the bluest clusters in dIs.
Finally, we include dwarf galaxies in the Local Group with old,
metal-poor GCs \citep[e.g.][]{dcmould88,mateo98}.  To convert from
galaxy luminosity to stellar mass we adopt $M/L_V = 5$ for all dE/dSph
galaxies and $M/L_V = 2$ for all dI galaxies.  To bridge the mass gap
between dwarf and giant elliptical galaxies and we calculate $T_{\rm
MP}$ for the \hst\ ACS Virgo Cluster sample of 100 early-type galaxies
\citep{cote04,peng06}.  The number of metal-poor clusters is
$(1-f_{red})N_{GC}$ from Table~2 of \cite{peng06}.  Galaxy masses are
calculated from $B_T$ given in \cite{cote04}.  As with the other
dwarfs, we assume $(B-V) = 0.77$ and $M/L_V = 5$.  Since the brightest
galaxies in the ACS sample are much larger than the ACS field-of-view,
the number of GCs given in \cite{peng06} will give systematically low
values of $T_{\rm MP}$ without additional corrections.  Therefore, we
plot only galaxies with $B_T>11$ ($M_V > -20.7$).  Also, for galaxies
with $M_V < -19.5$ we multiply the number of GCs by 1.5 to roughly
correct for the fact that the spatial extent of these GCS is somewhat
larger than the ACS field.  The final corrected numbers of GCs for the
ACS sample are being computed by the ACS Virgo team \citep{jordan06}.
Finally, for a comparison with spiral and giant elliptical galaxies we
use the values of $T_{\rm MP}$ from \cite{goud03}, \cite{chandar04},
and \cite{rz05}.

\begin{figure}
\includegraphics[width=\textwidth]{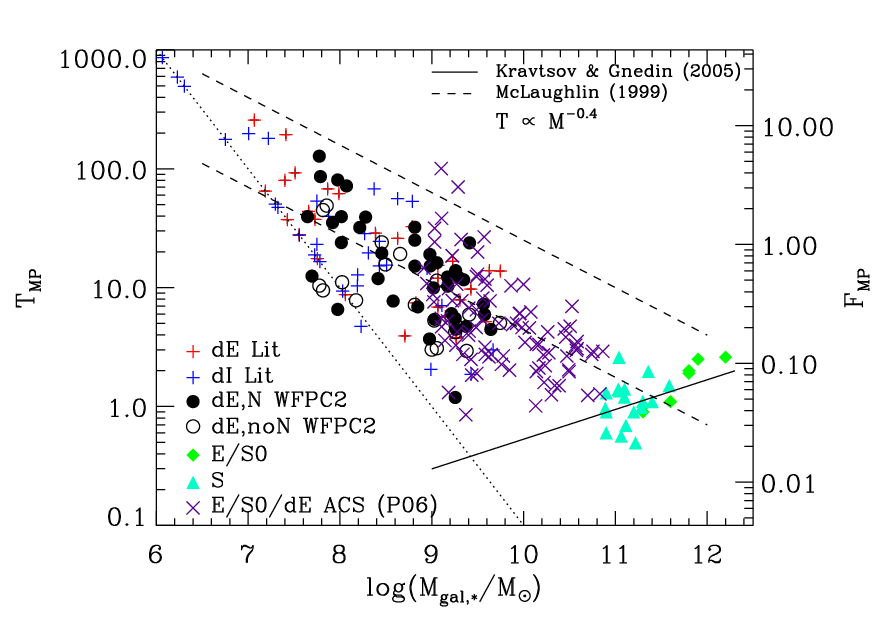}
\caption{The T parameter for metal-poor (MP) GC populations vs. galaxy
  stellar mass for dE galaxies from the current work and from the
  literature (see text for references).  The equivalent mass fraction,
  \fmp, is shown on the right axis assuming a universal GC mass
  function.  The dotted line on the left is the line of constant
  $N_{\rm GC} = 1$.  The solid line is the prediction of \fmp\ with
  galaxy mass for $\log(M_{G,*}) > 10.5$ from \cite{kravtsov05}.  The
  dashed lines have a slope of $-0.4$, from the SNe-driven wind models
  of \cite{mclaughlin99}.  The lower dashed line is the prediction
  from \cite{mclaughlin99} for $M/L_V=5$ for the galaxies.  The upper
  dashed line is an approximation of the upper envelope to the
  points.}
\label{fig:tflit}
\end{figure}

Figure~\ref{fig:tflit} shows how $T_{\rm MP}$ depends on galaxy
stellar mass.  On the right axis we also show the equivalent mass
fraction of GCs, $F_{\rm MP}$, assuming a universal GCLF so that
$F_{\rm MP} = 0.0433T_{\rm MP}$ (see above). It is notable that there
is a single trend of $T_{\rm MP}$ that is obeyed by all types of
galaxies over 6 orders of magnitude of galaxy mass.  The lower limit
to the distribution is the line of $N_{\rm GC} = 1$ (dotted line).  A
few points fall below this line because the statistical background
subtraction of the \hst\ survey can result in the net number of GCs
being less than one. There also appears to be an upper envelope to the
values of $T_{\rm MP}$ that is a line from the points for the faintest
dwarfs through the points for the most massive ellipticals.  The
equation of this line is
\begin{equation}
\log(T_{\rm MP}) = 5.4 - 0.4\log(M/M_{\odot}).
\label{eqn:tmax}
\end{equation}

Table~\ref{tab:meanprop} gives the mean values of \sn\, $T_{\rm MP}$,
and $F_{\rm MP}$ for different galaxy samples.  For the dwarfs the
statistics are computed for galaxies with $7.5 < \log(M/M_{\odot}) <
10$ to avoid biasing the results from the literature samples upward.
These statistics do not include the Virgo ACS sample because of the
many uncertain corrections that are needed to $N_{\rm GC}$. The mean
properties from the \hst\ dE snapshot survey agree well with the
values for dEs from the literature.  The importance of correcting for
galaxy $M/L$ using $T$ is also seen for the dIs.  The mean \sn\ for
the dIs is about a factor of 3 lower than the mean \sn\ for the dEs.
However, the mean $T_{\rm MP}$ and $F_{\rm MP}$ for the dIs is
comparable to that for the dEs, especially dE,noNs.  Giant galaxies
have much smaller values of mean \tmp\ and \fmp\ than in dwarfs.  As
shown in previous studies, giant cluster ellipticals do have somewhat
larger values of \tmp\ than field ellipticals or spirals \citep{rz05}.

However, the comparison between different samples of dEs and dIs is
complicated since the sample of dI galaxies in rich clusters that have
been surveyed for GCs is still very small and it is biased towards
more massive galaxies. Also, comparing mean properties is problematic
given the increase in dispersion with diminishing mass.

Therefore, we have also run two-side Kolmogorov-Smirnov tests to
compare the cumulative distributions of \tmp\ for different samples
(Table~\ref{tab:ksresults}).  In Table~\ref{tab:ksresults} a N, HST,
or noN designation refers to the \hst\ Snapshot Survey, ``Lit'' denotes
data taken from the literature, G is for galaxies in groups or the
field, ``C'' is for galaxies in the Virgo or Fornax Clusters.  As before,
we have not included the ACS Virgo galaxies in the statistics. This shows
that \tmp\ from the \hst\ Snapshot Survey and from the literature are
consistent with each other.  However, the distribution of \tmp\ in
dE,N and dE,noN galaxies are not consistent with being drawn from the
same distribution.  At low masses \tmp\ for dE,noN and dI galaxies are
very similar but the probabilities diminish somewhat for higher
galaxy masses.  The distribution of \tmp\ for dE,N galaxies is
similar to that for the small sample of dI galaxies in clusters, but
not to that of dIs in groups.  

Now we are in a position to try to understand these trends. That giant
elliptical galaxies have higher \sn\ or \tmp\ than spirals can be
explained by hierarchical merging \citep[see][]{rz05,bekki06b}.
\cite{kravtsov05} have run cosmological N-body simulations of the
formation of galaxies with halo masses greater than $10^{10}$~\Msun\
and found a correlation on GC mass fraction with halo mass.  Their
prediction of the baryonic mass fraction of GCs is the solid line in
Figure~\ref{fig:tflit}.  This prediction agrees with with the measured
values of \fmp\ for spiral and massive ellipticals.  This trend
results from a relatively constant fraction of halo mass being turned
into GCs (also see \cite{mclaughlin99}).

The trend of increasing \tmp\ or \fmp\ seen below $\log(M/M_{\sun})
\approx 10.5$ is thought to exist because lower mass halos are more
susceptible to losing gas via supernovae and stellar winds generated
by the formation of massive clusters.  This will reduce the average
star formation rate in lower-mass dwarfs and increase the relative
fraction of mass in massive star clusters, thus increasing \tmp.
\cite{mclaughlin99} used the SNe wind models of \cite{ds86} to
reproduce the trends in \sn\ seen by \cite{durrell96} and
\cite{mil98}. In this CDM-based model $T_{\rm MP} \propto M^{-0.4}$
and the two dashed lines in Figure~\ref{fig:tflit} show this trend
with two normalizations.  The upper line is Equation~\ref{eqn:tmax}
and defines the upper envelope of the \tmp\ trend.  The other line is
$\log(T_{\rm MP}) = 4.645 - 0.4\log(M/M_{\odot})$ and is the
prediction of Equation~27 from \cite{mclaughlin99} for $M/L_{V,GC}=2$,
$M/L_{V,Gal}=5$, and a GC formation efficiency of 0.26\%.  This
confirms that this model is a good representation of the data.

Recently \cite{forbes05} discussed the trend of increasing \sn\ with
decreasing mass using only the ACS Virgo data.  The increase in \sn\
for $\log(M/M_{\sun}) < 10.5$ is compared to new feedback models of
\cite{db04} which predict that $M/L_V \propto M^{-2/3}$ or $L_V
\propto M^{5/3}$.  As shown by \cite{forbes05}, $S_{\rm N} \propto
M^{-5/3}$ is a reasonable match to the ACS Virgo data assuming that
$N_{\rm GC}$ is constant.  However, $N_{\rm GC}$ is not constant with
galaxy mass and a slope this steep is does not match the trends of
\sn\ or \tm\ when all the galaxies from the current analysis are
considered.  However, it can be shown that the prediction of
\cite{db04} that $M/L_V \propto M^{-2/3}$ is consistent with
$M_{gas}^{init}/M_{*} \propto L_*^{-0.4}$ found by \cite{mclaughlin99}
using the \cite{ds86} wind models.  Therefore, the new wind models do
appear to be consistent with the previous calculations and with the
data.

\cite{bekki06b} have recently run N-body cosmological simulations to
investigate the origin of this trend.  They find that the data for
galaxies with $M_V < -16$ can be reproduced if $M/L_V \propto M^{-1}$
and if the formation of metal-poor GCs is truncated at $z=15$
(presumably due to winds or reionization).  The lower \sn\ or \tmp\ in
higher-mass dwarfs is due to the merging of small halos, many of which
do not have GCs since they virialized after $z=15$.  It would be
interesting to compare the results of these simulations with the
larger set of dwarf galaxy data presented here.  Also, observational
ages and total halo masses are needed to test the results of these
simulations. 

At a given mass there is considerable scatter in \tmp, especially for
the dwarfs, from 0 to a value given approximately by
Equation~\ref{eqn:tmax}.  At least some of this scatter should be due
to intrinsic variations in properties and processes.  In the model of
\cite{mclaughlin99} variations in \tmp\ would be produced by
variations in the GC formation efficiency, $\epsilon_{\rm cl} = M^{\rm
init}_{\rm gcs}/M^{\rm init}_{\rm gas}$.  Therefore, natural
variations in the initial gas fraction and the details of the star
formation and ISM-feedback processes will affect \tmp.  It is likely
that stochastic effects become more important in lower-mass galaxies
since single star formation events can have global impacts.
Variations in $M/L$ will also contribute to the scatter.

The differences between nucleated and non-nucleated dEs and the
comparison with dEs and dIs in clusters and the field suggest shows
that environmental effects are important.  The similarities of the GCS
properties between dE,noN galaxies in clusters and dI galaxies
suggests that dE,noNs may be the result of dI galaxies that were
simply stripped of gas when they fell into the cluster.  However, in
dE,N star and cluster formation occurred in addition to gas
stripping.  \cite{bekki06} has shown that interactions can produce gas
inflow and central star formation that can lead to nuclei.  Therefore,
it is likely that the progenitors of dE,N galaxies were those that
experienced significant interactions, either because they were formed
deeper in the potential well of the galaxy cluster or because of the
particular characteristics of their orbits.

Virgo and Fornax represent different environments in that they have
very different galaxy number densities, yet we don't see a significant
difference in the dE GCLFs of the two clusters.  Processes on a
``local scale'' that depend on the dwarf's location within a cluster
can affect its star cluster formation history hence the difference
between dE,N, dE,noN and dIs.  But these processes are fairly similar
in Virgo and Fornax (and don't strongly depend on the galaxy cluster
mass or density) so we don't encounter a large difference between
the two galaxy clusters.

Presumably the cumulative effect of ``local'' processes that affect
star cluster formation (e.g.\ interactions, ram-pressure stripping) do
depend on the galaxy cluster properties.  For example, cluster
velocity dispersion affects the nature of interactions and the
efficiency of ram pressure stripping. Therefore, more massive clusters
could have different dE to dI and dE,N to dE,noN ratios than less
massive clusters.  The similarities seen here between Virgo and Fornax
may be due to the selection effect that we are only considering true
dEs that have completed the transformation from dIs. Therefore, the
star-formation histories of dEs in Virgo and Fornax are not that
different. The GCLFs for a mass-selected dwarf galaxy population that
includes more dI and transition objects may show some dependence on galaxy
cluster properties.

Higher star formation rates will produce both new star clusters and
field stars.  To enhance \sn\ or \tmp\, the process must be biased in
favor of star clusters.  This would occur if star cluster formation
occurs first and the subsequent supernovae, perhaps aided by ram
pressure stripping, truncates field star formation
\citep{mclaughlin99}.  This is plausible since star cluster formation
probably occurs in the densest, most rapidly collapsing regions of
gas.  The field stars that are formed in the galaxies centers will
produce the higher central surface brightnesses seen in luminous dE,N
(see Figure~\ref{fig:snsb}). Luminous nuclei can then form via the
merging of central GCs via dynamical friction
\citep{hg98,ohlin00,bekki06,miocchi06}.  Because the star formation
occurs over an extended period, especially in the more massive
galaxies, this will lead to massive nuclei which are redder, younger,
and more metal-rich than the typical GC \citep[see][]{lmf04}.
Some predictions of simple dynamical friction calculations do not
agree with the observations \citep{lotz01} so additional observations
and more sophisticated modeling are needed.

\section{Conclusions}\label{sec:concl}

This paper has explored the globular cluster luminosity functions and
specific frequencies for the full sample of 69 galaxies in the \hst\
WFPC2 dE Snapshot Survey.  The data has been combined with new
complementary surveys of star clusters in a wide range of galaxy
types. The main results are:

\begin{enumerate}
\item The GCLFs of the dEs in the present sample are well-represented
  with a $t_5$ function.  The metallicity-corrected peak of the GCLF
  for Virgo dEs is $M_{V,Z}^0(dE,HST) = -7.3 \pm 0.1$.  This is
  consistent with most other recent measurements of the GCLF peaks in
  dEs. This result along with recent results from the literature give
  a fairly clear indication that the GCLF peak in dEs is fainter by
  $\sim0.3$ magnitudes, on average, than the metal-poor peak in
  spirals and giant ellipticals.  This is suggestive that less GC
  destruction has occurred in dwarfs but more data is needed to
  confirm this result.
\item The bright-end slope of the luminosity distribution has
  a power-law form with slope $\alpha = -1.9\pm0.1$, consistent with
  the bright-end slope in most previous studies and with the slope of
  the mass function of young, luminous clusters in starbursting
  systems.  
\item The trend of increasing \sn\ with decreasing luminosity seen in
  earlier studies is confirmed. Likewise, we continue to see that dE,N
  galaxies have a mean value of \sn\ about a factor of two higher than
  the mean for dE,noN galaxies.
\item In order to compare the metal-poor GCSs of different types of
  galaxies we compute \tmp, the number of GCs normalized to galaxy
  mass. All galaxies with $\log(M/M_{\odot}) < 10.5$ show a consistent
  trend of increasing \tmp\ with decreasing galaxy mass.  The slope of
  the trend in \tmp\ is consistent with the SNe-wind models of
  \cite{ds86} and \cite{mclaughlin99}.
\item Assuming a universal GCLF the percentage of stellar mass in GCs
  is $F_{\rm MP} = 0.0433T_{\rm MP}$.  For galaxies that have formed
  GCs, \fmp\ ranges from 0.02 for spirals and low-luminosity
  ellipticals to about 10 for the faintest dwarfs.
\end{enumerate}

Recent large, uniform surveys of star cluster systems in all types of
galaxies, dwarf galaxies in particular, are starting to reveal more
clearly the complicated interaction between cluster formation, the
feedback of that formation on the evolution of galaxies, and the
importance of environments. While the cluster mass function appears to
be universal within the uncertainties, there is growing evidence that
the bend in the GCLF of dwarf galaxies is at a lower luminosity, or
mass, than in giant galaxies due to less dynamical evolution. It is
still uncertain whether the GCLF of low mass galaxies is a nearly
pristine power law. Also, clearly feedback and galaxy mass are
fundamental parameters. Theories of supernovae driven winds can
explain the tends of \tmp\ with galaxy mass and show that low-mass
galaxies will have difficulty retaining the products of star
formation. This also explains why higher mass dwarfs have a higher
fraction of redder, more metal-rich star clusters
\citep{strader06,peng06}. However, the differences in \tmp\ and the
spatial distributions of dE,N and dE,noN galaxies also indicate the
importance of environment (i.e. cluster core, cluster outskirts, or
field).

While this work has illuminated some of the fundamental processes of
star cluster formation, (i.e. the shape of the cluster mass function
and the dependence of \tmp\ and \fmp\ on host galaxy mass), future
work is needed to constrain cluster formation histories and their
dependence on environment. The ages and metallicities either measured
directly or inferred from colors suggest a common period of star
cluster formation in all galaxy types and environments very soon after
the first stars began to form.  However, as explained in general terms
by theories and simulations of galaxy and structure formation, a given
potential well in a high density environment can collapse earlier and
have a higher average star formation rates than the same potential in
a low-density environments. This picture predicts that star clusters
formed in high-density environments should be slightly older than
those formed in low-density environments. This may be seen in globular
clusters in the Milky Way that are thought to have been accreted from
dwarf galaxies \citep{mackey04}. More accurate abundances,
$[\alpha/{\rm Fe}]$ ratios, and ages of a wide sample of star clusters
and nuclei are needed to confirm these predictions. Surveying star
clusters in dwarf galaxies is a demonstrated fruitful approach to
studying the formation and evolution of both star clusters and
galaxies.

\acknowledgments

We would like to thank the referee for useful suggestions and Dean
McLaughlin for useful discussions about this work. This research has
made use of NASA's Astrophysics Data System Bibliographic Services and
the NASA/IPAC Extragalactic Database (NED) which is operated by the
Jet Propulsion Laboratory, California Institute of Technology, under
contract with the National Aeronautics and Space Administration.  B.\
W.\ M.\ is supported by the Gemini Observatory, which is operated by
the Association of Universities for Research in Astronomy, Inc., on
behalf of the international Gemini partnership of Argentina,
Australia, Brazil, Canada, Chile, the United Kingdom, and the United
States of America. J.\ M.\ L.\ acknowledges support from the NOAO Leo
Goldberg Fellowship.


\begin{deluxetable}{lrrrrrrr}
\tablewidth{0pt}
\tablecaption{Basic Properties of Nucleated Galaxies in the \hst\ dE
  Snapshot Survey}
\tablehead{
\colhead{Galaxy} &
\colhead{Type\tablenotemark{a}} &
\colhead{RA} &
\colhead{Dec} &
\colhead{$B$\tablenotemark{a}} &
\colhead{A$_B$\tablenotemark{b}} &
\colhead{$M_B$\tablenotemark{c}} &
\colhead{$N_{\rm GC}$\tablenotemark{d}} \\
 & & \multicolumn{1}{c}{(J2000)} & \multicolumn{1}{c}{(J2000)} & & & & 
}
\startdata
\cola LGC0050\colb dE,N\colc  10:51:01.60\cold   13:20:01.0\cole 16.7\colf
0.06\colg $      -13.36$\colh $ 7.2\pm  3.8$\eol
\cola FCC0025\colb dE0,N\colc   3:23:33.40\cold  -36:59:02.0\cole 17.7\colf
0.00\colg $    -13.70$\colh $ 2.5\pm  3.7$\eol
\cola FCC0046\colb dE4\colc   3:26:25.02\cold  -37:07:41.4\cole 15.6\colf
0.00\colg $      -15.80$\colh $ 8.0\pm  4.2$\eol
\cola FCC0059\colb dE0,N\colc   3:27:46.60\cold  -33:33:51.0\cole 19.4\colf
0.00\colg $    -12.00$\colh $ 0.0\pm  3.2$\eol
\cola FCC0136\colb dE2,N\colc   3:34:29.48\cold  -35:32:47.0\cole 14.8\colf
0.00\colg $    -16.60$\colh $18.0\pm  5.3$\eol
\cola FCC0146\colb dE4,N\colc   3:35:11.60\cold  -35:19:23.0\cole 19.5\colf
0.00\colg $    -11.90$\colh $ 0.5\pm  2.7$\eol
\cola FCC0150\colb dE4,N\colc   3:35:24.09\cold  -36:21:49.6\cole 15.7\colf
0.00\colg $    -15.70$\colh $ 8.0\pm  4.2$\eol
\cola FCC0174\colb dE1,N\colc   3:36:45.49\cold  -33:00:50.5\cole 16.7\colf
0.00\colg $    -14.70$\colh $ 8.0\pm  3.7$\eol
\cola FCC0189\colb dE4,N\colc   3:37:08.21\cold  -34:43:54.3\cole 18.8\colf
0.00\colg $    -12.60$\colh $ 0.5\pm  3.4$\eol
\cola FCC0238\colb dE5,N\colc   3:40:17.19\cold  -36:32:05.5\cole 18.7\colf
0.00\colg $    -12.70$\colh $ 2.0\pm  2.0$\eol
\cola FCC0242\colb   dE5\colc   3:40:20.54\cold  -37:38:41.0\cole 17.8\colf
0.00\colg $     -13.60$\colh $ 0.0\pm  2.7$\eol
\cola FCC0246\colb  dE2\colc   3:40:37.70\cold  -36:07:16.0\cole 19.1\colf
0.00\colg $     -12.30$\colh $ 0.0\pm  2.7$\eol
\cola FCC0254\colb dE0,N\colc   3:41:00.80\cold  -35:44:33.0\cole 17.6\colf
0.00\colg $   -13.80$\colh $ 4.5\pm  4.2$\eol
\cola FCC0316\colb dE3,N\colc   3:47:01.40\cold  -36:26:15.0\cole 16.7\colf
0.00\colg $   -14.70$\colh $17.0\pm  5.6$\eol
\cola FCC0324\colb dS01(8)\colc   3:47:52.67\cold  -36:28:18.1\cole 15.3\colf
0.00\colg $ -16.10$\colh $ 9.0\pm  4.6$\eol
\cola VCC0158\colb dE3,N\colc  12:15:40.07\cold   15:00:20.1\cole 15.8\colf
0.11\colg $   -15.21$\colh $ 8.9\pm  4.6$\eol
\cola VCC0240\colb dE2,N\colc  12:17:31.33\cold   14:21:20.4\cole 18.2\colf
0.14\colg $   -12.84$\colh $ 7.2\pm  4.1$\eol
\cola VCC0452\colb dE4,N\colc  12:21:04.74\cold   11:45:17.6\cole 15.8\colf
0.01\colg $   -15.11$\colh $15.5\pm  5.3$\eol
\cola VCC0503\colb dE3,N\colc  12:21:50.24\cold    8:32:27.8\cole 16.8\colf
0.01\colg $   -14.11$\colh $ 2.5\pm  3.7$\eol
\cola VCC0529\colb dE4,N\colc  12:22:08.59\cold    9:53:39.9\cole 18.2\colf
0.00\colg $   -12.70$\colh $ 3.5\pm  3.5$\eol
\cola VCC0646\colb dE3\colc  12:23:31.80\cold   17:47:40.5\cole18.8\colf
0.03\colg $     -12.13$\colh $ 4.5\pm  3.9$\eol
\cola VCC0747\colb dE0,N\colc  12:24:47.77\cold    8:59:29.2\cole 16.2\colf
0.00\colg $  -14.70$\colh $14.0\pm  4.9$\eol
\cola VCC0871\colb dE4,N\colc  12:26:05.63\cold   12:33:33.7\cole 15.4\colf
0.09\colg $  -15.59$\colh $15.6\pm  4.9$\eol
\cola VCC0896\colb dE3,N\colc  12:26:22.48\cold   12:46:59.9\cole 17.8\colf
0.10\colg $  -13.20$\colh $ 4.5\pm  3.9$\eol
\cola VCC0940\colb dE1,N\colc  12:26:47.05\cold   12:27:14.6\cole 14.8\colf
0.09\colg $  -16.19$\colh $52.4\pm  8.1$\eol
\cola VCC0949\colb dE4,N\colc  12:26:54.55\cold   10:39:57.1\cole 15.1\colf
0.00\colg $  -15.80$\colh $19.7\pm  5.8$\eol
\cola VCC0965\colb dE7,N\colc  12:27:03.30\cold   12:33:37.8\cole 15.4\colf
0.10\colg $  -15.60$\colh $13.2\pm  4.6$\eol
\cola VCC0992\colb dE0,N\colc  12:27:18.63\cold    8:12:45.5\cole 15.8\colf
0.00\colg $  -15.10$\colh $ 3.0\pm  4.1$\eol
\cola VCC1073\colb dE3,N\colc  12:28:08.60\cold   12:05:35.7\cole 14.2\colf
0.07\colg $  -16.77$\colh $16.7\pm  5.8$\eol
\cola VCC1077\colb dE0,N\colc  12:28:10.36\cold   12:48:24.7\cole 19.2\colf
0.08\colg $  -11.78$\colh $ 1.5\pm  4.5$\eol
\cola VCC1105\colb dE0,N\colc  12:28:27.24\cold   14:09:20.9\cole 16.2\colf
0.08\colg $  -14.78$\colh $ 4.1\pm  3.4$\eol
\cola VCC1107\colb dE4,N\colc  12:28:30.55\cold    7:19:29.7\cole 15.1\colf
0.01\colg $  -15.81$\colh $21.6\pm  5.5$\eol
\cola VCC1252\colb dE0,N\colc  12:30:01.74\cold    9:28:25.7\cole 18.8\colf
0.00\colg $  -12.10$\colh $ 6.5\pm  4.2$\eol
\cola VCC1254\colb dE0,N\colc  12:30:05.09\cold    8:04:23.6\cole 15.0\colf
0.00\colg $  -15.90$\colh $20.5\pm  8.0$\eol
\cola VCC1272\colb dE1,N\colc  12:30:15.54\cold   13:18:25.9\cole 18.5\colf
0.06\colg $  -12.46$\colh $ 2.5\pm  3.4$\eol
\cola VCC1308\colb dE6,N\colc  12:30:45.95\cold   11:20:35.5\cole 15.1\colf
0.02\colg $  -15.82$\colh $ 7.0\pm  4.3$\eol
\cola VCC1311\colb dE1,N\colc  12:30:47.32\cold    7:36:19.1\cole 15.6\colf
0.00\colg $  -15.30$\colh $15.7\pm  4.7$\eol
\cola VCC1363\colb dE3,N\colc  12:31:27.84\cold   10:55:50.5\cole 19.0\colf
0.00\colg $  -11.90$\colh $ 0.0\pm  2.5$\eol
\cola VCC1386\colb dE3,N\colc  12:31:51.40\cold   12:39:26.0\cole 14.4\colf
0.07\colg $  -16.57$\colh $22.7\pm  6.1$\eol
\cola VCC1497\colb dE4,N\colc  12:33:18.53\cold   17:27:32.7\cole 15.7\colf
0.04\colg $  -15.24$\colh $ 5.0\pm  4.6$\eol
\cola VCC1514\colb dE7,N\colc  12:33:37.72\cold    7:52:16.5\cole 15.1\colf
0.00\colg $  -15.80$\colh $ 1.8\pm  2.5$\eol
\cola VCC1530\colb dE2,N\colc  12:33:55.69\cold    5:43:07.6\cole 18.3\colf
0.00\colg $  -12.60$\colh $ 6.5\pm  4.2$\eol
\cola VCC1577\colb dE4\colc  12:34:38.37\cold   15:36:09.7\cole 15.8\colf
0.02\colg $    -15.12$\colh $12.5\pm  5.0$\eol
\cola VCC1714\colb dE4,N\colc  12:37:25.65\cold   14:18:48.3\cole 18.5\colf
0.03\colg $  -12.43$\colh $ 0.0\pm  3.8$\eol
\cola VCC1876\colb dE5,N\colc  12:41:20.42\cold   14:42:01.9\cole 14.9\colf
0.03\colg $  -16.03$\colh $22.1\pm  6.3$\eol
\enddata
\label{tab:densmpl}
\tablenotetext{a}{Morphological types and apparent $B$ magnitudes
  taken from the following catalogs: FCC = Fornax Cluster Catalog
  \citep{f89}; VCC = Virgo Cluster Catalog \citep{bst85}; LGC =
  Leo Group Catalog \citep{fs90}.}
\tablenotetext{b}{Foreground extinction from \cite{schlegel98} as
  reported in the NASA Extragalactic Database (NED).}
\tablenotetext{c}{Absolute $B$ magnitudes, corrected for foreground
  extinction,  calculated assuming
  $(m-M)_0 = 30.0$ for the Leo Group, 30.92 for the Virgo Cluster, and
  31.39 for the Fornax Cluster \citep{freedman01}.}
\tablenotetext{d}{Corrected for background contamination.  From
  \cite{lmf04} with galaxy association corrected.}
\end{deluxetable}

\begin{deluxetable}{lrrrrrrr}
\tablewidth{0pt}
\tablecaption{Basic Properties of Non-nucleated Galaxies in the \hst\ dE
  Snapshot Survey}
\tablehead{
\colhead{Galaxy} &
\colhead{Type\tablenotemark{a}} &
\colhead{RA} &
\colhead{Dec} &
\colhead{$B$\tablenotemark{a}} &
\colhead{A$_B$\tablenotemark{b}} &
\colhead{$M_B$\tablenotemark{c}} &
\colhead{$N_{\rm GC}$\tablenotemark{d}} \\
 & & \multicolumn{1}{c}{(J2000)} & \multicolumn{1}{c}{(J2000)} & & & &
}
\startdata
\cola LGC0047\colb dE\colc  10:50:18.97\cold   13:16:21.1\cole 14.9\colf
0.06\colg $       -15.16$\colh $ 2.9\pm  4.4$\eol
\cola FCC0027\colb dE2\colc   3:23:55.76\cold  -34:48:26.5\cole 19.3\colf
0.00\colg $     -12.10$\colh $ 0.5\pm  3.1$\eol
\cola FCC0048\colb dE3\colc   3:26:42.55\cold  -34:32:49.9\cole 17.1\colf
0.00\colg $     -14.30$\colh $ 7.0\pm  4.4$\eol
\cola FCC0064\colb dE5\colc   3:28:00.49\cold  -38:32:15.1\cole 17.5\colf
0.00\colg $     -13.90$\colh $ 0.0\pm  2.9$\eol
\cola FCC0110\colb dE4\colc   3:32:57.37\cold  -35:44:15.5\cole 16.8\colf
0.00\colg $     -14.60$\colh $ 0.0\pm  4.2$\eol
\cola FCC0144\colb dE0\colc   3:35:00.30\cold  -35:19:20.0\cole 19.2\colf
0.00\colg $     -12.20$\colh $ 0.5\pm  3.4$\eol
\cola FCC0212\colb dE1?\colc   3:38:20.90\cold  -36:24:42.0\cole 17.6\colf
0.00\colg $    -13.80$\colh $ 5.5\pm  5.3$\eol
\cola FCC0218\colb dE4\colc   3:38:45.40\cold  -35:15:59.0\cole 18.5\colf
0.00\colg $     -12.90$\colh $ 0.0\pm  2.7$\eol
\cola FCC0304\colb dE1\colc   3:45:30.90\cold  -34:30:18.0\cole 18.8\colf
0.00\colg $     -12.60$\colh $ 0.0\pm  3.1$\eol
\cola VCC0009\colb dE1,N\colc  12:09:22.34\cold   13:59:33.1\cole 13.9\colf
0.02\colg $   -17.02$\colh $23.9\pm  6.9$\eol
\cola VCC0118\colb dE3\colc  12:14:36.85\cold    9:41:21.8\cole 15.6\colf
0.00\colg $     -15.30$\colh $ 3.0\pm  3.6$\eol
\cola VCC0128\colb dE0\colc  12:14:59.51\cold    9:33:55.0\cole 15.6\colf
0.00\colg $     -15.30$\colh $11.1\pm  4.3$\eol
\cola VCC0543\colb dE5\colc  12:22:19.53\cold   14:45:38.3\cole 14.8\colf
0.08\colg $     -16.18$\colh $12.9\pm  4.4$\eol
\cola VCC0546\colb dE6\colc  12:22:21.59\cold   10:36:07.1\cole 15.7\colf
0.03\colg $     -15.23$\colh $ 4.8\pm  3.3$\eol
\cola VCC0917\colb dE6\colc  12:26:32.37\cold   13:34:43.5\cole 14.9\colf
0.11\colg $     -16.11$\colh $ 6.0\pm  5.1$\eol
\cola VCC0996\colb dE5\colc  12:27:21.17\cold   13:06:36.3\cole 18.4\colf
0.09\colg $     -12.59$\colh $ 0.0\pm  3.7$\eol
\cola VCC1651\colb dE5\colc  12:36:07.46\cold    6:03:17.4\cole 17.0\colf
0.00\colg $      -13.90$\colh $ 4.2\pm  5.3$\eol
\cola VCC1729\colb dE5?\colc  12:37:46.06\cold   10:59:07.1\cole 17.8\colf
0.00\colg $     -13.10$\colh $ 1.0\pm  3.0$\eol
\cola VCC1762\colb dE6\colc  12:38:32.18\cold   10:22:35.6\cole 16.2\colf
0.00\colg $      -14.70$\colh $ 4.0\pm  3.5$\eol
\cola VCC1781\colb dE4\colc  12:39:11.77\cold    8:04:25.5\cole 18.7\colf
0.00\colg $      -12.20$\colh $ 2.5\pm  3.4$\eol
\cola VCC1877\colb dE2\colc  12:41:24.22\cold    8:21:57.1\cole 18.6\colf
0.00\colg $      -12.30$\colh $ 3.0\pm  3.3$\eol
\cola VCC1948\colb dE3\colc  12:42:58.02\cold   10:40:54.5\cole 15.1\colf
0.00\colg $      -15.80$\colh $ 6.5\pm  3.7$\eol
\cola VCC2008\colb dE5\colc  12:44:47.48\cold   12:03:51.5\cole 15.1\colf
0.01\colg $      -15.81$\colh $ 7.5\pm  5.1$\eol
\cola VCC2029\colb dE3\colc  12:45:40.60\cold    9:24:18.6\cole 18.2\colf
0.02\colg $      -12.72$\colh $ 1.0\pm  3.6$\eol
\enddata
\label{tab:denonsmpl}
\tablenotetext{a}{Morphological types and apparent $B$ magnitudes 
  taken from the following catalogs: FCC = Fornax Cluster Catalog
  \citep{f89}; VCC = Virgo Cluster Catalog \citep{bst85}; LGC =
  Leo Group Catalog \citep{fs90}.}
\tablenotetext{b}{Foreground extinction from \cite{schlegel98} as
  reported in the NASA Extragalactic Database (NED).}
\tablenotetext{c}{Absolute $B$ magnitudes, corrected for foreground
  extinction, calculated assuming
  $(m-M)_0 = 30.0$ for the Leo Group, 30.92 for the Virgo Cluster, and
  31.39 for the Fornax Cluster \citep{freedman01}.}
\tablenotetext{d}{Corrected for background contamination. From
  \cite{lmf04} with galaxy association corrected.}
\end{deluxetable}

\begin{deluxetable}{crcrc}
\tablewidth{0pt}

\tablecaption{$V$-band GCLF for all Virgo galaxies}
\tablehead{
\colhead{$V$} &
\colhead{N$_{\rm GC + nuc}$} &
\colhead{$f_{\rm bkg,GC + nuc}$} &
\colhead{N$_{\rm GC}$} &
\colhead{$f_{\rm bkg,GC}$} \\
\colhead{(1)} &
\colhead{(2)} &
\colhead{(3)} &
\colhead{(4)} &
\colhead{(5)} 
}
\startdata
 17.666&$   2.1\pm1.4$ & 0.90 & $ 2.0\pm1.4$ &  1.00\\
 17.998&$   3.0\pm1.7$ & 0.89 & $ 2.9\pm1.7$ &  0.89\\
 18.330&$   2.1\pm1.4$ & 0.78 & $ 2.0\pm1.4$ &  0.89\\
 18.662&$   2.1\pm1.4$ & 0.71 & $ 0.9\pm0.9$ &  0.83\\
 18.994&$   4.0\pm2.0$ & 0.40 & $ 4.0\pm2.0$ &  0.50\\
 19.326&$   3.0\pm1.7$ & 0.50 & $ 2.9\pm1.7$ &  0.50\\
 19.658&$   7.0\pm2.6$ & 0.74 & $ 6.0\pm2.5$ &  0.79\\
 19.990&$   4.9\pm2.2$ & 0.50 & $ 4.9\pm2.2$ &  0.54\\
 20.322&$  10.0\pm3.2$ & 0.48 & $10.1\pm3.2$ &  0.57\\
 20.654&$   7.0\pm2.6$ & 0.42 & $ 6.0\pm2.5$ &  0.50\\
 20.986&$   7.9\pm2.8$ & 0.43 & $ 6.9\pm2.6$ &  0.50\\
 21.318&$   9.1\pm3.0$ & 0.21 & $ 6.0\pm2.5$ &  0.26\\
 21.650&$  13.0\pm3.6$ & 0.36 & $12.1\pm3.5$ &  0.40\\
 21.982&$  22.1\pm4.7$ & 0.22 & $18.1\pm4.3$ &  0.26\\
 22.314&$  36.1\pm6.0$ & 0.20 & $33.1\pm5.8$ &  0.22\\
 22.646&$  44.9\pm6.7$ & 0.21 & $42.9\pm6.5$ &  0.23\\
 22.978&$  68.0\pm8.2$ & 0.19 & $63.9\pm8.0$ &  0.20\\
 23.310&$  74.0\pm8.6$ & 0.23 & $71.1\pm8.4$ &  0.23\\
 23.642&$  81.9\pm9.1$ & 0.27 & $81.1\pm9.0$ &  0.27\\
 23.974&$  81.9\pm9.1$ & 0.30 & $80.0\pm8.9$ &  0.31\\
 24.306&$  91.9\pm9.6$ & 0.48 & $90.9\pm9.5$ &  0.49\\
 24.638&$  74.9\pm8.7$ & 0.64 & $75.1\pm8.7$ &  0.65\\
 24.970&$  46.1\pm6.8$ & 0.73 & $46.0\pm6.8$ &  0.75\\
 25.302&$   4.0\pm2.0$ & 0.73 & $ 4.0\pm2.0$ &  0.77
\enddata
\label{tab:gclfvgo}
\tablecomments{(1) $V$ magnitude of bin center; (2) Total GC$+$nuclei
    in bin (histogram in Figure~\ref{fig:gclfallv}a),
    $\phi(V)\times701\times0.332$ with Poisson errors; (3) Background
    fraction of objects in (2); (4) Like (2) but GC candidates only,
    $\phi(V)\times673\times0.332$; (5) Background fraction of objects
    in (4).}
\end{deluxetable}

\begin{deluxetable}{crcrc}
\tablewidth{0pt}
\tablecaption{$V$-band GCLF for all Fornax galaxies}
\tablehead{
\colhead{$V$} &
\colhead{N$_{\rm GC + nuc}$} &
\colhead{$f_{\rm bkg,GC + nuc}$} &
\colhead{N$_{\rm GC}$} &
\colhead{$f_{\rm bkg,GC}$} \\
\colhead{(1)} &
\colhead{(2)} &
\colhead{(3)} &
\colhead{(4)} &
\colhead{(5)} 
}
\startdata
 17.666 & $ 0.0\pm0.0$ & 1.00 & $0.0\pm0.0$ & 1.00\\
 17.998 & $ 2.0\pm1.4$ & 0.94 & $2.0\pm1.4$ & 1.00\\
 18.330 & $ 0.0\pm0.0$ & 0.92 & $0.0\pm0.0$ & 1.00\\
 18.662 & $ 0.0\pm0.0$ & 0.89 & $0.0\pm0.0$ & 1.00\\
 18.994 & $ 0.0\pm0.0$ & 0.80 & $0.0\pm0.0$ & 1.00\\
 19.326 & $ 2.0\pm1.4$ & 0.83 & $1.0\pm1.0$ & 1.00\\
 19.658 & $ 1.0\pm1.0$ & 0.95 & $1.0\pm1.0$ & 0.97\\
 19.990 & $ 2.0\pm1.4$ & 0.85 & $1.0\pm1.0$ & 1.00\\
 20.322 & $ 2.0\pm1.4$ & 0.85 & $1.0\pm1.0$ & 0.99\\
 20.654 & $ 1.0\pm1.0$ & 0.84 & $0.0\pm0.0$ & 0.96\\
 20.986 & $ 1.0\pm1.0$ & 0.83 & $1.0\pm1.0$ & 0.94\\
 21.318 & $ 2.0\pm1.4$ & 0.64 & $2.0\pm1.4$ & 0.90\\
 21.650 & $ 4.0\pm2.0$ & 0.78 & $2.0\pm1.4$ & 0.93\\
 21.982 & $ 4.0\pm2.0$ & 0.62 & $2.0\pm1.4$ & 0.84\\
 22.314 & $ 4.0\pm2.0$ & 0.57 & $3.0\pm1.7$ & 0.76\\
 22.646 & $ 9.0\pm3.0$ & 0.56 & $7.0\pm2.7$ & 0.71\\
 22.978 & $ 9.0\pm3.0$ & 0.50 & $9.0\pm3.0$ & 0.59\\
 23.310 & $16.0\pm4.0$ & 0.51 & $15.0\pm3.9$ & 0.54\\
 23.642 & $21.0\pm4.6$ & 0.52 & $21.0\pm4.6$ & 0.51\\
 23.974 & $29.0\pm5.4$ & 0.51 & $27.0\pm5.2$ & 0.49\\
 24.306 & $24.0\pm4.9$ & 0.65 & $24.0\pm4.9$ & 0.66\\
 24.638 & $31.0\pm5.6$ & 0.74 & $31.0\pm5.6$ & 0.80\\
 24.970 & $25.0\pm5.0$ & 0.75 & $25.0\pm5.0$ & 0.85\\
 25.302 & $ 9.0\pm3.0$ & 0.68 & $9.0\pm3.0$ & 0.84
\enddata
\label{tab:gclffnx}
\tablecomments{(1) $V$ magnitude of bin center; (2) Total GC$+$nuclei
    in bin (histogram in Figure~\ref{fig:gclfallv}b),
    $\phi(V)\times198\times0.332$; (3) Background fraction of objects
    in (2); (4) Like (2) but GC candidates only,
    $\phi(V)\times184\times0.332$; (5) Background fraction of objects
    in (4).}
\end{deluxetable}

\begin{deluxetable}{crcrc}
\tablewidth{0pt}
\tablecaption{I-band GCLF (GC + nuclei) for Virgo and Fornax galaxies}
\tablehead{
\colhead{$I$} &
\colhead{N$_{\rm Virgo}$} &
\colhead{$f_{\rm bkg,Virgo}$} &
\colhead{N$_{\rm Fornax}$} &
\colhead{$f_{\rm bkg,Fornax}$} \\
\colhead{(1)} &
\colhead{(2)} &
\colhead{(3)} &
\colhead{(4)} &
\colhead{(5)} 
}
\startdata
 16.686&$ 3.1\pm1.8$ &  1.00 &$ 1.0\pm 1.0$ &  0.96\\
 17.058&$ 2.1\pm1.4$ &  1.00 &$ 1.0\pm 1.0$ &  1.00\\
 17.430&$ 1.0\pm1.0$ &  0.90 &$ 0.0\pm 0.0$ &  0.94\\
 17.802&$ 3.1\pm1.8$ &  0.95 &$ 0.0\pm 0.0$ &  0.89\\
 18.174&$ 2.1\pm1.4$ &  0.73 &$ 0.0\pm 0.0$ &  1.00\\
 18.546&$ 8.1\pm2.8$ &  0.81 &$ 2.0\pm 1.4$ &  0.96\\
 18.918&$ 6.0\pm2.4$ &  0.70 &$ 2.0\pm 1.4$ &  0.86\\
 19.290&$ 7.0\pm2.7$ &  0.72 &$ 2.0\pm 1.4$ &  0.92\\
 19.662&$12.0\pm3.5$ &  0.50 &$ 1.0\pm 1.0$ &  0.84\\
 20.034&$ 8.1\pm2.8$ &  0.56 &$ 3.0\pm 1.7$ &  0.86\\
 20.406&$14.1\pm3.8$ &  0.40 &$ 1.0\pm 1.0$ &  0.81\\
 20.778&$15.1\pm3.9$ &  0.29 &$ 8.0\pm 2.8$ &  0.72\\
 21.150&$27.9\pm5.3$ &  0.26 &$ 3.0\pm 1.7$ &  0.69\\
 21.522&$42.0\pm6.5$ &  0.15 &$ 5.0\pm 2.2$ &  0.53\\
 21.894&$63.1\pm7.9$ &  0.20 &$11.0\pm 3.3$ &  0.60\\
 22.266&$74.1\pm8.6$ &  0.23 &$13.0\pm 3.6$ &  0.59\\
 22.638&$98.1\pm9.9$ &  0.27 &$21.0\pm 4.6$ &  0.56\\
 23.010&$99.9\pm0.0$ &  0.33 &$32.0\pm 5.7$ &  0.52\\
 23.382&$93.1\pm9.6$ &  0.53 &$33.0\pm 5.7$ &  0.62\\
 23.754&$92.1\pm9.6$ &  0.68 &$31.0\pm 5.6$ &  0.69\\
 24.126&$30.0\pm5.5$ &  0.73 &$24.0\pm 4.9$ &  0.79\\
 24.498&$ 0.0\pm0.0$ &  0.19 &$ 4.0\pm 2.0$ &  0.46\\
\enddata
\label{tab:gclfi}
\tablecomments{(1) $I$ magnitude of bin center; (2) Total Virgo
    GC$+$nuclei in bin, $\phi(I)\times701\times0.372$ (histogram in
    Figure~\ref{fig:gclfalli}a); (3) Background fraction of objects in
    (2); (4) Like (2) but for Fornax galaxies,
    $\phi(V)\times198\times0.372$ (histogram in
    Figure~\ref{fig:gclfalli}b); (5) Background fraction of objects in
    (4).}
\end{deluxetable}

\begin{deluxetable}{lcccc}
\tablewidth{0pt}
\tablecaption{Best-fitting Parameters of dE GCLFs}
\tablehead{
\colhead{Sample} &
\colhead{$m^0$} &
\colhead{$\sigma_t$\tablenotemark{a}} &
\colhead{$M^0$\tablenotemark{b}}
}
\startdata
\sidehead{$V$-band}
Virgo (GC$+$nuclei) & $23.61 \pm 0.11$ & $1.08 \pm 0.09$ & $-7.31\pm0.11$\\
Virgo (GC only) & $23.63 \pm 0.09$ & $1.02 \pm 0.09$  & $-7.29\pm0.10$\\
Virgo (dE,N, GC only) & $23.69 \pm 0.11$ & $1.06 \pm 0.10$  & $-7.23\pm0.11$\\
Virgo (dE,noN, GC only) & $23.42 \pm 0.17$ & $0.77 \pm 0.18$  & $-7.50\pm0.17$\\
Virgo bright (GC$+$nuc) & $23.64 \pm 0.09$ & $0.97 \pm 0.08$  & $-7.28\pm0.09$\\
Virgo faint (GC$+$nuc) & $23.18 \pm 0.43$ & $1.63 \pm 0.30$  & $-7.74\pm0.43$\\

Fornax (GC$+$nuclei) & $24.15 \pm 0.40$ & $1.03 \pm 0.30$  & $-7.24\pm0.40$\\
Fornax (GC only) & $24.00 \pm 0.28$ & $0.67 \pm 0.23$  & $-7.39\pm0.30$\\
Nuclei & $21.26\pm0.21$ & $1.11\pm0.18$  & $-9.7\pm0.2$  \\
\sidehead{$I$-band}
Virgo (GC$+$nuclei) & $22.80 \pm 0.11$ & $1.1 \pm 0.1$  & $-8.12\pm0.11$\\
Fornax (GC$+$nuclei) & $23.40 \pm 0.40$ & $1.0 \pm 0.3$ & $-8.00\pm0.40$
\enddata
\label{tab:gclf}
\tablenotetext{a}{Equivalent Gaussian sigma: $\sigma_G \approx 1.29\sigma_t$.}
\tablenotetext{b}{Assuming $(m-M)_{0}^{Virgo} = 30.92$, $(m-M)_{0}^{Fornax} = 31.39$}
\end{deluxetable}

\begin{deluxetable}{ccc}
\tablewidth{0pt}
\tablecaption{GCLF Parameter Simulations}
\tablehead{
\colhead{Galaxy $\sigma_D$} &
\colhead{$V^0$\tablenotemark{a}} &
\colhead{$\sigma_t$\tablenotemark{b}}
}
\startdata
1.0 Mpc & $23.59 \pm 0.09$ & $1.11 \pm 0.05$\\
2.0 Mpc & $23.59 \pm 0.08$ & $1.13 \pm 0.04$\\
3.0 Mpc & $23.56 \pm 0.08$ & $1.18 \pm 0.03$\\
4.0 Mpc & $23.52 \pm 0.12$ & $1.23 \pm 0.06$\\
\enddata
\label{tab:gclfsim}
\tablenotetext{a}{Input value is 23.62}
\tablenotetext{b}{Input value is 1.1}
\end{deluxetable}

\begin{deluxetable}{crcrc}
\tablewidth{0pt}
\tablecaption{V-band GCLF for Virgo galaxies with $M_V > -15.75$}
\tablehead{
\colhead{$V$} &
\colhead{N$_{\rm GC + nuc}$} &
\colhead{$f_{\rm bkg,GC + nuc}$} &
\colhead{N$_{\rm GC}$} &
\colhead{$f_{\rm bkg,GC}$} \\
\colhead{(1)} &
\colhead{(2)} &
\colhead{(3)} &
\colhead{(4)} &
\colhead{(5)} 
}
\startdata
 17.666&$   0.0\pm0.0$ & 0.83 & $ 0.0\pm0.0$ &  0.80\\
 17.998&$   0.0\pm0.0$ & 0.76 & $ 0.0\pm0.0$ &  0.77\\
 18.330&$   0.0\pm0.0$ & 0.71 & $ 0.0\pm0.0$ &  0.68\\
 18.662&$   0.0\pm0.0$ & 0.54 & $ 0.0\pm0.0$ &  0.55\\
 18.994&$   4.0\pm2.0$ & 0.32 & $ 4.0\pm2.0$ &  0.31\\
 19.326&$   0.0\pm0.0$ & 0.29 & $ 0.0\pm0.0$ &  0.30\\
 19.658&$   2.0\pm1.4$ & 0.63 & $ 2.0\pm1.4$ &  0.62\\
 19.990&$   2.0\pm1.4$ & 0.37 & $ 2.0\pm1.4$ &  0.37\\
 20.322&$   3.0\pm1.7$ & 0.40 & $ 3.0\pm1.7$ &  0.44\\
 20.654&$   2.0\pm1.4$ & 0.38 & $ 2.0\pm1.4$ &  0.42\\
 20.986&$   4.0\pm2.0$ & 0.42 & $ 4.0\pm2.0$ &  0.47\\
 21.318&$   7.0\pm2.6$ & 0.24 & $ 4.0\pm2.0$ &  0.29\\
 21.650&$   6.0\pm2.5$ & 0.43 & $ 5.0\pm2.2$ &  0.51\\
 21.982&$   9.0\pm3.0$ & 0.32 & $ 6.0\pm2.4$ &  0.39\\
 22.314&$   4.0\pm2.0$ & 0.34 & $ 3.0\pm1.7$ &  0.42\\
 22.646&$   8.0\pm2.8$ & 0.41 & $ 7.0\pm2.6$ &  0.49\\
 22.978&$  16.0\pm4.0$ & 0.43 & $14.0\pm3.7$ &  0.50\\
 23.310&$  14.0\pm3.7$ & 0.52 & $12.0\pm3.5$ &  0.58\\
 23.642&$  21.0\pm4.6$ & 0.60 & $21.0\pm4.6$ &  0.65\\
 23.974&$  18.0\pm4.2$ & 0.64 & $18.0\pm4.2$ &  0.68\\
 24.306&$  24.0\pm4.9$ & 0.78 & $24.0\pm4.9$ &  0.80\\
 24.638&$  19.0\pm4.4$ & 0.86 & $19.0\pm4.4$ &  0.86\\
 24.970&$  17.0\pm4.1$ & 0.90 & $17.0\pm4.1$ &  0.89\\
 25.302&$   3.0\pm1.7$ & 0.89 & $ 3.0\pm1.7$ &  0.88
\enddata
\label{tab:gclfvgo_faint}
\tablecomments{(1) $V$ magnitude of bin center; (2) Total GC$+$nuclei
    in bin, $\phi(V)\times183\times0.332$ (histogram in
    Figure~\ref{fig:gclf2}b); (3) Background fraction of objects in
    (2); (4) Like (2) but GC candidates only,
    $\phi(V)\times170\times0.332$; (5) Background fraction of objects
    in (4).}
\end{deluxetable}

\begin{deluxetable}{lrrrrr}
\tablewidth{0pt}
\tablecaption{Derived GCS Properties for dE,N in the \hst\ dE
  Snapshot Survey}
\tablehead{
\colhead{Galaxy} &
\colhead{$M_V$} &
\colhead{$N_{\rm GC}({\rm tot})$} &
\colhead{\sn} &
\colhead{\tmp} &
\colhead{\fmp}  
}
\startdata
\cola LGC0050\colb -14.13& $  7.5\pm  4.0$\cold$ 16.6\pm  8.9$\colf$ 39.2\pm 
21.0 $\colg$1.1\pm  0.9$\eol
\cola FCC0025\colb -14.47& $  3.1\pm  4.5$\cold$  5.0\pm  7.4$\colf$ 11.9\pm 
17.5 $\colg$0.7\pm  0.2$\eol
\cola FCC0046\colb -16.57& $  9.9\pm  5.3$\cold$  2.3\pm  1.2$\colf$  5.5\pm  
2.9 $\colg$0.2\pm  0.1$\eol
\cola FCC0059\colb -12.77& $  0.0\pm  4.0$\cold   $0.\pm 31.2$\colf$  0.0 \pm
73.7$ \colg$0.0\pm  1.7$\eol
\cola FCC0136\colb -17.37& $ 22.3\pm  6.6$\cold$  2.5\pm  0.7$\colf$  5.9\pm  
1.7 $\colg$0.2\pm  0.1$\eol
\cola FCC0146\colb -12.67& $  0.6\pm  3.4$\cold$  5.3\pm 29.0$\colf$ 12.5\pm 
68.5 $\colg$0.5\pm  0.0$\eol
\cola FCC0150\colb -16.47& $  9.9\pm  5.3$\cold$  2.6\pm  1.4$\colf$  6.0\pm  
3.2 $\colg$0.3\pm  0.0$\eol
\cola FCC0174\colb -15.47& $  9.9\pm  4.6$\cold$  6.4\pm  3.0$\colf$ 15.2\pm  
7.1 $\colg$0.4\pm  0.4$\eol
\cola FCC0189\colb -13.37& $  0.6\pm  4.2$\cold$  2.8\pm 18.9$\colf$  6.6\pm 
44.5 $\colg$0.8\pm  0.8$\eol
\cola FCC0238\colb -13.47& $  2.5\pm  2.5$\cold$ 10.1\pm 10.1$\colf$ 23.9\pm 
23.9 $\colg$0.9\pm  0.1$\eol
\cola FCC0242\colb -14.37& $  0.0\pm  3.3$\cold$  0.0\pm  5.9$\colf$  0.0\pm 
13.9 $\colg$0.0\pm  0.2$\eol
\cola FCC0246\colb -13.07& $  0.0\pm  3.4$\cold$  0.0\pm 20.1$\colf$  0.0\pm 
47.4 $\colg$0.0\pm  0.8$\eol
\cola FCC0254\colb -14.57& $  5.6\pm  5.2$\cold$  8.3\pm  7.7$\colf$ 19.5\pm 
18.2 $\colg$0.6\pm  0.3$\eol
\cola FCC0316\colb -15.47& $ 21.1\pm  6.9$\cold$ 13.7\pm  4.5$\colf$ 32.3\pm 
10.5 $\colg$0.9\pm  0.7$\eol
\cola FCC0324\colb -16.87& $ 11.2\pm  5.7$\cold$  2.0\pm  1.0$\colf$  4.7\pm  
2.4 $\colg$0.1\pm  0.1$\eol
\cola VCC0158\colb -15.98& $ 10.4\pm  5.4$\cold$  4.2\pm  2.2$\colf$ 10.0\pm  
5.1 $\colg$0.4\pm  0.1$\eol
\cola VCC0240\colb -13.61& $  8.5\pm  4.8$\cold$ 30.4\pm 17.2$\colf$ 71.8\pm 
40.7 $\colg$2.8\pm  0.4$\eol
\cola VCC0452\colb -15.88& $ 18.1\pm  6.2$\cold$  8.1\pm  2.8$\colf$ 19.1\pm  
6.6 $\colg$0.6\pm  0.4$\eol
\cola VCC0503\colb -14.88& $  2.9\pm  4.3$\cold$  3.3\pm  4.8$\colf$  7.7\pm 
11.3 $\colg$0.5\pm  0.2$\eol
\cola VCC0529\colb -13.47& $  4.1\pm  4.1$\cold$ 16.8\pm 16.9$\colf$ 39.6\pm 
39.9 $\colg$2.9\pm  1.7$\eol
\cola VCC0646\colb  -12.9& $  5.3\pm  4.6$\cold$ 36.5\pm 31.9$\colf$ 86.1\pm 
75.3 $\colg$4.8\pm  1.6$\eol
\cola VCC0747\colb -15.47& $ 16.4\pm  5.7$\cold$ 10.6\pm  3.7$\colf$ 25.1\pm  
8.8 $\colg$0.8\pm  0.4$\eol
\cola VCC0871\colb -16.36& $ 18.2\pm  5.8$\cold$  5.2\pm  1.7$\colf$ 12.3\pm  
3.9 $\colg$0.4\pm  0.2$\eol
\cola VCC0896\colb -13.97& $  5.3\pm  4.6$\cold$ 13.6\pm 11.9$\colf$ 32.1\pm 
28.1 $\colg$1.8\pm  0.5$\eol
\cola VCC0940\colb -16.96& $ 61.4\pm  9.5$\cold$ 10.1\pm  1.6$\colf$ 23.8\pm  
3.7 $\colg$0.8\pm  0.3$\eol
\cola VCC0949\colb -16.57& $ 23.1\pm  6.9$\cold$  5.4\pm  1.6$\colf$ 12.8\pm  
3.8 $\colg$0.4\pm  0.2$\eol
\cola VCC0965\colb -16.37& $ 15.4\pm  5.3$\cold$  4.4\pm  1.5$\colf$ 10.3\pm  
3.6 $\colg$0.6\pm  0.2$\eol
\cola VCC0992\colb -15.87& $  3.5\pm  4.8$\cold$  1.6\pm  2.2$\colf$  3.7\pm  
5.1 $\colg$0.3\pm  0.2$\eol
\cola VCC1073\colb -17.54& $ 19.6\pm  6.8$\cold$  1.9\pm  0.7$\colf$  4.4\pm  
1.5 $\colg$0.2\pm  0.0$\eol
\cola VCC1077\colb -12.55& $  1.8\pm  5.3$\cold$ 16.8\pm 50.7$\colf$ 39.6\pm
119.6 $\colg$2.6\pm  1.3$\eol
\cola VCC1105\colb -15.55& $  4.8\pm  4.0$\cold$  2.9\pm  2.4$\colf$  6.9\pm  
5.7 $\colg$0.6\pm  0.4$\eol
\cola VCC1107\colb -16.58& $ 25.3\pm  6.5$\cold$  5.9\pm  1.5$\colf$ 13.9\pm  
3.6 $\colg$0.5\pm  0.2$\eol
\cola VCC1252\colb -12.87& $  7.6\pm  4.9$\cold$ 54.1\pm 34.8$\colf$127.8\pm 
82.2 $\colg$5.0\pm  0.8$\eol
\cola VCC1254\colb -16.67& $ 24.0\pm  9.3$\cold$  5.2\pm  2.0$\colf$ 12.2\pm  
4.7 $\colg$0.7\pm  0.2$\eol
\cola VCC1272\colb -13.23& $  2.9\pm  4.0$\cold$ 14.9\pm 20.3$\colf$ 35.3\pm 
47.8 $\colg$1.6\pm  0.1$\eol
\cola VCC1308\colb -16.59& $  8.2\pm  5.1$\cold$  1.9\pm  1.2$\colf$  4.5\pm  
2.8 $\colg$0.3\pm  0.2$\eol
\cola VCC1311\colb -16.07& $ 18.4\pm  5.5$\cold$  6.9\pm  2.1$\colf$ 16.2\pm  
4.9 $\colg$0.7\pm  0.0$\eol
\cola VCC1363\colb -12.67& $  0.0\pm  3.0$\cold$  0.0\pm 25.5$\colf$  0.0\pm 
60.3 $\colg$0.0\pm  0.4$\eol
\cola VCC1386\colb -17.34& $ 26.6\pm  7.2$\cold$  3.1\pm  0.8$\colf$  7.3\pm  
2.0 $\colg$0.3\pm  0.0$\eol
\cola VCC1497\colb -16.01& $  5.8\pm  5.4$\cold$  2.3\pm  2.1$\colf$  5.4\pm  
5.0 $\colg$0.2\pm  0.1$\eol
\cola VCC1514\colb -16.57& $  2.1\pm  2.9$\cold$  0.5\pm  0.7$\colf$  1.2\pm  
1.6 $\colg$0.1\pm  0.0$\eol
\cola VCC1530\colb -13.37& $  7.6\pm  4.9$\cold$ 34.2\pm 22.0$\colf$ 80.6\pm 
51.8 $\colg$3.1\pm  0.5$\eol
\cola VCC1577\colb -15.89& $ 14.6\pm  5.9$\cold$  6.4\pm  2.6$\colf$ 15.2\pm  
6.1 $\colg$0.6\pm  0.2$\eol
\cola VCC1714\colb  -13.2& $  0.0\pm  4.5$\cold$  0.0\pm 23.5$\colf$  0.0\pm 
55.6 $\colg$0.0\pm  0.3$\eol
\cola VCC1876\colb  -16.8& $ 25.9\pm  7.4$\cold$  4.9\pm  1.4$\colf$ 11.7\pm  
3.3 $\colg$0.4\pm  0.2$\eol
\enddata
\label{tab:denprop}
\end{deluxetable}

\begin{deluxetable}{lrrrrr}
\tablewidth{0pt}
\tablecaption{Derived GCS Properties for dE,noN in
  the \hst\ dE Snapshot Survey}
\tablehead{
\colhead{Galaxy} &
\colhead{$M_V$} &
\colhead{$N_{\rm GC}({\rm tot})$} &
\colhead{\sn} &
\colhead{\tmp} &
\colhead{\fmp} 
}
\startdata
\cola LGC0047\colb -15.93& $  3.0\pm  4.5$\cold$  1.3\pm  1.9$\colf$  3.0\pm  
4.5 $\colg$0.1\pm  0.0$\eol
\cola FCC0027\colb -12.87& $  0.6\pm  3.8$\cold$  4.4\pm 27.1$\colf$ 10.4\pm 
64.1 $\colg$1.5\pm  1.5$\eol
\cola FCC0048\colb -15.07& $  8.7\pm  5.4$\cold$  8.1\pm  5.1$\colf$ 19.2\pm 
12.0 $\colg$0.5\pm  0.5$\eol
\cola FCC0064\colb -14.67& $  0.0\pm  3.6$\cold$  0.0\pm  4.9$\colf$  0.0\pm 
11.6 $\colg$0.0\pm  0.3$\eol
\cola FCC0110\colb -15.37& $  0.0\pm  5.2$\cold$  0.0\pm  3.7$\colf$  0.0\pm  
8.7 $\colg$0.0\pm  0.4$\eol
\cola FCC0144\colb -12.97& $  0.6\pm  4.2$\cold$  4.0\pm 27.2$\colf$  9.5\pm 
64.3 $\colg$0.7\pm  0.4$\eol
\cola FCC0212\colb -14.57& $  6.9\pm  6.6$\cold$ 10.2\pm  9.8$\colf$ 24.0\pm 
23.2 $\colg$1.2\pm  0.2$\eol
\cola FCC0218\colb -13.67& $  0.0\pm  3.4$\cold$  0.0\pm 11.6$\colf$  0.0\pm 
27.3 $\colg$0.0\pm  0.5$\eol
\cola FCC0304\colb -13.37& $  0.0\pm  3.8$\cold$  0.0\pm 17.1$\colf$  0.0\pm 
40.4 $\colg$0.0\pm  0.6$\eol
\cola VCC0009\colb -17.79& $ 28.0\pm  8.1$\cold$  2.1\pm  0.6$\colf$  5.1\pm  
1.5 $\colg$0.2\pm  0.0$\eol
\cola VCC0118\colb -16.07& $  3.5\pm  4.2$\cold$  1.3\pm  1.6$\colf$  3.1\pm  
3.7 $\colg$0.1\pm  0.0$\eol
\cola VCC0128\colb -16.07& $ 13.0\pm  5.1$\cold$  4.9\pm  1.9$\colf$ 11.5\pm  
4.5 $\colg$0.8\pm  0.5$\eol
\cola VCC0543\colb -16.95& $ 15.2\pm  5.1$\cold$  2.5\pm  0.8$\colf$  5.9\pm  
2.0 $\colg$0.2\pm  0.0$\eol
\cola VCC0546\colb   -16.& $  5.6\pm  3.9$\cold$  2.2\pm  1.6$\colf$  5.3\pm  
3.7 $\colg$0.2\pm  0.1$\eol
\cola VCC0917\colb -16.88& $  7.0\pm  6.0$\cold$  1.2\pm  1.1$\colf$  2.9\pm  
2.5 $\colg$0.2\pm  0.1$\eol
\cola VCC0996\colb -13.36& $  0.0\pm  4.4$\cold$  0.0\pm 20.0$\colf$  0.0\pm 
40.0 $\colg$0.0\pm  0.6$\eol
\cola VCC1651\colb -14.67& $  4.9\pm  6.2$\cold$  6.6\pm  8.4$\colf$ 15.7\pm 
19.7 $\colg$0.5\pm  0.2$\eol
\cola VCC1729\colb -13.87& $  1.2\pm  3.5$\cold$  3.3\pm  9.9$\colf$  7.8\pm 
23.5 $\colg$0.6\pm  0.3$\eol
\cola VCC1762\colb -15.47& $  4.7\pm  4.1$\cold$  3.0\pm  2.6$\colf$  7.2\pm  
6.2 $\colg$0.2\pm  0.1$\eol
\cola VCC1781\colb -12.97& $  2.9\pm  4.0$\cold$ 19.1\pm 25.7$\colf$ 45.0\pm 
60.8 $\colg$7.5\pm  7.8$\eol
\cola VCC1877\colb -13.07& $  3.5\pm  3.9$\cold$ 20.8\pm 23.0$\colf$ 49.1\pm 
54.3 $\colg$1.8\pm  0.4$\eol
\cola VCC1948\colb -16.57& $  7.6\pm  4.3$\cold$  1.8\pm  1.0$\colf$  4.2\pm  
2.4 $\colg$0.2\pm  0.0$\eol
\cola VCC2008\colb -16.58& $  8.8\pm  6.0$\cold$  2.1\pm  1.4$\colf$  4.9\pm  
3.3 $\colg$0.2\pm  0.0$\eol
\cola VCC2029\colb -13.49& $  1.2\pm  4.2$\cold$  4.7\pm 17.0$\colf$ 11.1\pm 
40.1 $\colg$0.4\pm  0.1$\eol
\enddata
\label{tab:denonprop}
\end{deluxetable}

\begin{deluxetable}{lcccc}
\tablewidth{0pt}
\tablecaption{Mean GCS Properties for Different Galaxy Types}
\tablehead{
\colhead{Sample} &
\colhead{\sn} &
\colhead{$T_{\rm MP}$} &
\colhead{$F_{\rm MP}$} &
\colhead{$\langle\log(M_{gal,*})\rangle$} 
}
\startdata
dE,N WFPC2      & 8.7  &   20.4 &   0.9 & 8.66\eol
dE,noN WFPC2    & 4.3  &   10.2 &   0.7 & 8.60\eol
dE WFPC2        & 7.2  &   16.9 &   0.8 & 8.64\eol
dE literature   & 8.6  &   21.0 &   0.9 & 8.65\eol
dE all          & 7.6  &   17.9 &   0.8 & 8.64\eol
dE cluster      & 3.7  &   16.3 &   0.7 & 9.22\eol
dE group        &11.3  &   26.8 &   1.2 & 8.14\eol
                &      &        &  &  \eol
dI literature   & 2.6  &   15.0 &   0.7 & 8.23\eol
dI cluster      & 3.7  &   21.2 &   0.9 & 8.48\eol
dI group        & 2.0  &   12.0 &   0.5 & 8.12\eol
                &      &        &  &  \eol
E/S0            & 2-10 &   1.7  &  0.08 & 11.41\eol
Spirals         & $<2$ &   1.2  &  0.05 & 11.17\eol
\enddata
\label{tab:meanprop}
\end{deluxetable}

\begin{deluxetable}{lrrr}
\tablewidth{0pt}
\tablecaption{K-S Test Probabilities for \tmp\ in different ranges in log(M)} 
\tablehead{
\colhead{Sample} &
\colhead{P(7.5-8.5)} &
\colhead{P(8.5-9.8)} &
\colhead{P(7.5-9.8)} 
}
\startdata
dE(N) vs dE(noN)     &  0.26   &   0.03   &   0.03\eol
dE(HST) vs dE(lit)   &  0.47   &   0.96   &   0.71\eol
dE(C) vs dI(Lit)     &  0.34   &   0.16   &   0.03\eol
dE(G) vs dI(Lit)     &  0.04   &   0.91   &   0.27\eol
dE(C) vs dE(G)       &  0.47   &   0.07   &   0.15\eol
dE(C) vs dI(C)       &  0.93   &   0.19   &   0.27\eol
dI(C) vs dI(G)       &  0.35   &   0.05   &   0.03\eol
dE(noN) vs dI(C)     &  0.93   &   0.03   &   0.09\eol
dE(noN) vs dI(G)     &  0.92   &   0.07   &   0.11\eol
dE(noN) vs dI(Lit)   &  0.93   &   0.42   &   0.30\eol
dE(N) vs dI(G)       &  0.25   &   0.01   & $<0.01$\eol
dE(N) vs dI(C)       &  0.54   &   0.29   &   0.64\eol
dE(N) vs dI(Lit)     &  0.09   &   0.17   &   0.05\eol
\enddata
\label{tab:ksresults}
\end{deluxetable}


\end{document}